\documentclass[10pt,a4paper]{article}
\usepackage{inputenc}
\usepackage{setspace}
\usepackage{amsmath}
\usepackage{amsfonts}
\usepackage{amssymb}
\usepackage{amsthm}
\usepackage{graphicx}
\usepackage{enumitem}
\graphicspath{ {./figures/} }

\usepackage[margin=1in]{geometry}
\usepackage[title]{appendix}
\usepackage{authblk}
\usepackage{xcolor}
\usepackage{times}
\usepackage{natbib}
\usepackage{graphicx}
\usepackage{float}
\usepackage{xcolor}
\usepackage{amsfonts}
\usepackage{booktabs} 
\usepackage{caption}
\usepackage{subcaption}
\usepackage{enumitem}
\usepackage{mathrsfs} 
\usepackage{accents}
\usepackage{xr}
\usepackage{hyperref}
\usepackage{optidef}
\usepackage[mathscr]{euscript}
\usepackage{bbm}
\usepackage{adjustbox}
\usepackage{multirow}

\theoremstyle{plain}
\newtheorem{theorem}{Theorem}[section]
\newtheorem{proposition}[theorem]{Proposition}

\newtheorem{corollary}[theorem]{Corollary}
\theoremstyle{definition}

\newtheorem{assumption}[theorem]{Assumption}
\theoremstyle{remark}
\newtheorem{remark}[theorem]{Remark}

\DeclareMathOperator*{\argmin}{argmin}
\newcommand{\calA}{\mathcal{A}}
\newcommand{\calP}{\mathcal{P}}
\newcommand{\bh}{\boldsymbol{h}}
\newcommand{\bX}{\boldsymbol{X}}
\newcommand{\bO}{\boldsymbol{O}}
\newcommand{\bx}{\boldsymbol{x}}
\newcommand{\bm}{\boldsymbol{m}}

\newcommand{\bU}{\boldsymbol{U}}
\newcommand{\bOne}{\boldsymbol{1}}
\newcommand{\EE}{\mathbb{E}}
\newcommand{\PP}{\mathbb{P}}
\newcommand{\bphi}{\boldsymbol{\phi}}
\newcommand{\bPhi}{\boldsymbol{\Phi}}
\newcommand{\bpsi}{\boldsymbol{\psi}}
\newcommand{\bPsi}{\boldsymbol{\Psi}}
\newcommand{\btheta}{\boldsymbol{\theta}}

\usepackage{caption}
\usepackage[noend]{algpseudocode}

\usepackage{tikz}
\usetikzlibrary{positioning}

\tikzset{
    every neuron/.style={
        circle,
        draw,
        minimum size=1cm
    },
    neuron missing/.style={
        draw=none, 
        scale=4,
        text height=0.333cm,
        execute at begin node=\color{black}$\vdots$
    },
}

\usepackage{natbib}
\title{Efficient Inference for Covariate-adjusted Bradley-Terry Model with Covariate Shift}
\author[$\dagger$]{Xiudi Li$^\star$}
\author[$\ddagger$]{Sijia Li$^\star$}
\affil[$\dagger$]{Division of Biostatistics, University of California, Berkeley}
\affil[$\ddagger$]{Department of Biostatistics, Harvard T.H. Chan School of Public Health }

\date{April 8th, 2025}

\begin{document}
\maketitle
\def\thefootnote{$\star$}\footnotetext{These authors contributed equally to this work}

\begin{abstract}
  We propose a general framework for statistical inference on the overall strengths of players in pairwise comparisons, allowing for potential shifts in the covariate distribution. These covariates capture important contextual information that may impact the winning probability of each player. We measure the overall strengths of players under a target distribution through its Kullback-Leibler projection onto a class of covariate-adjusted Bradley-Terry model. Consequently, our estimands remain well-defined without requiring stringent model assumptions. We develop semiparametric efficient estimators and corresponding inferential procedures that allow for flexible estimation of the nuisance functions. When the assumptions of the covariate-adjusted Bradley-Terry model hold, we propose additional estimators that do not require observing all pairwise comparisons. We demonstrate the performance of our proposed method in simulation studies and apply it to assess the alignment of large language models with human preferences in real-world applications.
\end{abstract}

\section{Introduction}
\label{sec:intro}
The Bradley-Terry (BT) model \citep{bradley1952rank} is a widely used probability framework for pairwise comparisons, and forms the basis for ranking systems such as the Elo rating system \citep{elo1967proposed}. It is widely applied in various areas, including sports, genetics, psychology \citep{turner2012bradley}, and more recently, evaluation of large language models (LLMs) \citep{chiang2024chatbot}. In the classical BT model, each player is associated with a strength parameter $\theta$, and the win probability equals to the sigmoid transformation of the difference between the strengths of the two players in contest. However, as it is parametric, the BT model is subject to misspecification \citep{tang2025elo}. Moreover, it implicitly assumes transitivity -- if player $i$ is more likely to win against player $j$ and player $j$ is more likely to win against player $k$, then $i$ is also more likely to win against $k$. Such transitivity may not hold in practical applications \citep{spearing2023modeling}, and whether human preferences are transitive remains an active area of research \citep{regenwetter2011transitivity,birnbaum2023testing}.

There have been various extensions of the BT model that incorporate covariate information on either the players or the contests. Existing frequentist approaches, such as \cite{francis2010modeling}, \cite{turner2012bradley}, and \cite{schauberger2019btllasso}, as well as Bayesian approaches including \cite{seymour2022bayesian} and \cite{li2022bayesian}, typically assume that the strength is a linear combination of covariates. However, such linearity assumptions may be violated in practice. As such, \cite{baldassarre2023bradley} proposed a regression tree based approach to account for potential interactions among covariates. Relatedly, dynamic BT models aim at incorporating temporal information, and can be estimated using smoothing techniques \citep{cattelan2013dynamic} or spectral methods \citep{tian2024spectral}. 

Statistical inference in the BT model framework has also been studied extensively in the literature. Bradley \citep{bradley1955rank} established the asymptotic normality of the MLE in the classical BT model (see also \cite{simons1999asymptotics}) when the number of contests goes to infinity for a given set of players. As the model can be re-written as a logistic model, similar asymptotic normality results are expected when incorporating covariates under linearity assumptions (see, for example, \cite{springall1973response}.) This is a reasonable setup especially in contexts such as evaluating large language models, given the limited number of LLMs of interest and the large number of prompts they are evaluated on. We primarily work under this setting in this paper. However, we move beyond parametric BT model assumptions and consider pairwise comparisons nonparametrically accounting for covariate information, and tackle the problem of covariate shift. 

Another line of research considers the asymptotic regime where the number of contests between a pair of players remains fixed but the number of players tends to infinity. The seminal work \cite{simons1999asymptotics} established the asymptotic normality of the maximum likelihood estimates (MLE) under this setting when all pairwise comparisons are observed. Recent works along this direction extended such results and quantified the uncertainty in BT model estimation under a sparse comparison graph where only a small subset of pairwise comparisons are actually observed \citep{han2020asymptotic,gao2023uncertainty}. The recent work \cite{fan2024uncertainty} built upon this approach and incorporated covariate information under a linearity assumption. Finally, \cite{wang2024ranking} studied an extension of the BT model where the strength of each player is a nonparametric function of covariates and conducted inference based on multiplier bootstrap. Besides the different setting, all these methods assume that the BT assumption holds in that the pairwise winning probabilities are parameterized by the intrinsic strengths (conditioned on covariates or marginally) of each player.

Extending beyond parametric BT assumptions while incorporating covariate information is useful in many applications. One such application is in evaluating the alignment of LLMs with human preferences, which plays a critical role in ensuring accuracy and safety in the real-world deployment of these models. Recently developed platforms such as Chatbot Arena \citep{chiang2024chatbot} provide new opportunities to collect data on human preferences over LLM outputs in a cost-effective way via crowdsourcing and to evaluate LLMs over a diverse array of prompts expected in practical applications. In \cite{chiang2024chatbot}, the classical BT model was used to estimate the strength of each model and derive a ranking list. Such an approach measures the overall strengths of the models given the set of prompts in the dataset. When deployed for specific tasks, however, the distribution of the characteristics of the prompts may differ substantially from the benchmark dataset where pairwise comparison outcomes are annotated. We refer to such a scenario as one with \emph{covariate shift}. For example, a LLM-based AI math tutor \citep{bastani2024generative} may see more prompts and questions focused on math, reasoning, and general STEM knowledge. It is therefore of interest to evaluate LLMs under a shift in the characteristics in the prompts that can more accurately reflect real-world application for specific tasks.

Besides model evaluation and ranking, the BT model is also frequently used as a reward model in a typical Reinforcement Learning from Human Feedback (RLHF) workflow \citep{dong2024rlhf} which aims at incorporating human feedback into the training and fine-tuning of LLMs \citep{christiano2017deep}. Here the strength of each model is given by the reward function, a function of covariates on the prompts and the responses. From this perspective, inference on the BT model under potential covariate shift can be regarded as inference on the average reward achieved by a model in specific tasks. Beyond LLMs, covariate-adjusted BT model can be broadly used to understand the effects of specific attributes of the contests or the subjects making choices, in various application areas including sports and psychometrics. 

In this work, we develop semiparametric efficient estimators and the corresponding inferential procedures for the overall strength in the framework of Bradley-Terry model incorporating covariate information. Our main contributions are as follows. Firstly, we consider a conditional BT model where the strength of each player is a flexible function of the covariates. We study the conditional BT distribution that best approximates the target data generating mechanism, which is well-defined even if the BT model is misspecified. We show that under misspecification, the marginal strength of players can depend on the covariate distribution as well as the conditional probability each pair is sampled. Second, we develop semiparametric efficient estimators and construct confidence intervals of the overall strength that allow flexible estimation of the relevant nuisance functions via data-adaptive statistical learning tools. We also develop separate estimators that do not require observing all pairwise comparisons when the conditional BT model is correctly specified. Finally, we extend these inferential methods to situations with covariate shifts. 

\section{Problem formulation and estimand of interest}\label{sec:model and estimand}

Suppose that $K$ players are in the battle (e.g., in the context of evaluating LLMs, each LLM is a player.) Let $\calA = \{(k,l): 1\leq k < l \leq K\}$ denote the set of unordered pairs of indices. Let $\bX \in \mathcal{X} \subseteq \mathbb{R}^d$ denote a vector of covariates capturing the contextual information in each comparison, for example, nature of the prompt. In each comparison $i$, we observe the contextual information $\bX_i$, and a pair $A_i$ of players is drawn from the set $\calA$. We then observe the outcome of the comparison $Y_i \in \{0,1\}$ where $Y_i = 1$ indicates that the first player wins and $Y_i=0$ otherwise. Let $P$ denote the distribution of a generic data unit $(\bX,A,Y)$ in this labeled dataset, and let $P_{Y|(A,\bX)}$, $P_{A|\bX}$, and $P_{\bX}$ denote the conditional distribution of $Y$ given $(A,\bX)$, the conditional distribution of $A$ given $\bX$ and the marginal distribution of $\bX$, respectively, under sampling from $P$. Furthermore, let $Q$ denote a target distribution of interest, with $Q_{Y|(A,\bX)}$, $Q_{A|\bX}$, and $Q_{\bX}$ denoting the corresponding conditional and marginal distributions.

\subsection{Conditional Bradley-Terry model and Kullback-Leibler (KL) projection} 
The classical BT model assumes that $P(Y=1|A=(k,l)) = \sigma(\theta_k - \theta_l)$ where $\sigma(\cdot)$ is the sigmoid function and $\theta_k$ and $\theta_l$ are unknown parameters measuring the strengths of players $k$ and $l$. Hereafter, we refer to this model as the \emph{marginal} BT model.

In this paper, we consider a nonparametric extension of the BT model incorporating the contextual information. Furthermore, we use this nonparametric BT model as a potentially misspecified working model, and focus on the best approximation in terms of KL divergence of a target data generating distribution within this working model. Specifically, consider a \emph{working} model $\mathcal{P} = \{P_\theta: \btheta=(\theta_1,\ldots,\theta_K), \theta_k \in L^2(\mu)\}$ where $\mu$ is a dominating measure for $P$. For each $P_\theta \in \mathcal{P}$, the conditional distribution of $Y$ takes the form
\begin{equation}\label{eq: nonparametric BT}
    P_{Y|(A,\bX),\theta}(Y=1|A=(k,l),\bX=\bx) = \sigma(\theta_k(\bx) - \theta_l(\bx)),
\end{equation}
and $P_{A|\bX,\theta}$ and $P_{\bX,\theta}$ are unrestricted. We will refer to the conditional model containing distributions in \eqref{eq: nonparametric BT} as the \emph{conditional} BT model. In general, $\theta_k(\bx)$ is only identifiable up to a shift $c(\bx)$ that is independent of $k$. To resolve this issue, we impose the restriction that $\theta_l(\bx) = 0$ for all $\bx$ for some reference player $l$, and without loss of generality we take $l=1$. For any distribution $Q$, even those outside of $\mathcal{P}$, we can define its projection onto $\mathcal{P}$. Let $\rho_{kl}(\bx) = Q_{A|\bX}(A=(k,l)|\bX=\bx)$ and $q_{kl}(\bx) = Q_{Y|(A,\bX)}(Y=1|A=(k,l),\bX=\bx)$, for $1 \leq k < l \leq K$; and let $\rho_{lk}(\bx) = \rho_{kl}(\bx)$ and $q_{lk}(\bx) = 1 - q_{kl}(\bx)$, for $1 \leq k < l \leq K$.

\begin{proposition}[Kullback-Leibler projection of a target distribution]\label{prop: KL projection}
\sloppy Let $P_{\theta^*} = \argmin_{P_\theta \in \mathcal{P}} \EE_Q[\log(dQ/dP_\theta)]$. Then, for $Q_{\bX}$-almost every $\bx$ and all $k=2,\ldots,K$,
\begin{equation}\label{eq: EE given x}
    \sum_{l \neq k} \rho_{kl}(\bx)\left\{\sigma\left(\theta_k^*(\bx) - \theta_l^*(\bx)\right) - q_{kl}(\bx)\right\} = 0.
\end{equation}
\end{proposition}
The distribution $P_{\theta^*}$ achieves the smallest KL divergence to the target distribution, among all distributions in the conditional BT model. The indexing functions $\{\theta^*_k\}$ is implicitly defined pointwise via \eqref{eq: EE given x}, which involves the conditional probabilities $q_{kl}$. Notably, $\theta_k^*$ agrees with $\theta_{k,0}$ if $q_{kl}$ indeed takes the form in \eqref{eq: nonparametric BT} for some $\theta_{k,0}$ and $\theta_{l,0}$, and remains well defined even if the working model is misspecified. The projection $\theta^*_k$ may also depend on the probability $\rho$ under misspecification. Intuitively, if player $k$ is very strong against some player $l$ and is only matched up against player $l$ under $Q$, then player $k$ will appear strong overall and rank high. Therefore, to have a fair comparison between players, it is crucial to consider an appropriate choice of $Q_{A|\bX}$. In what follows, we will define our estimands under specific choice of $\rho$.

\subsection{Estimand of interest and identification}

\paragraph{Estimand of interest.} The value $\theta_k^*(\bx)$ quantifies the strength of player $k$ under a specific context $\bx$. We further summarize the overall strengths of the players by integrating over all contexts. Specifically, let $\btheta^*(\bx) = (\theta_2^*(\bx),\ldots,\theta_K^*(\bx))$ and $\bphi = \EE_{Q_{\bX}}[\btheta^*(\bX)] \in \mathbb{R}^{K-1}$ (note that $\theta_1^*(\bx) = 0$). We will consider statistical inference on $\bphi$ in situations with and without covariate shift (more details later.) 

In defining the estimand $\bphi$, we first project $Q_{Y|(A,\bX)}$ onto the conditional BT model, approximating the conditional probabilities $\{q_{kl}(\bx): k\neq l\}$ using $\{\theta^*_k(\bx):k\}$ pointwise, and then marginalize $\btheta^*(\bx)$ over the distribution of contextual information. Alternatively, one may average the winning probabilities $q_{kl}(\bx)$ over $\bx$ first to obtain a marginal winning probabilities, and then project onto the marginal BT model. Define $q_{kl} = E_{Q_{\bX}}[q_{kl}(\bX)]$ and $\rho_{kl} = E_{Q_{\bX}}[\rho_{kl}(\bX)]$. An alternative estimand of interest is $\bpsi = (\psi_2,\ldots,\psi_K) \in \mathbb{R}^{K-1}$ such that
\begin{equation}\label{eq: marginal EE}
    \sum_{l \neq k} \rho_{kl}\left\{\sigma\left(\psi_k - \psi_l\right) - q_{kl}\right\} = 0,
\end{equation}
where we again set $\psi_1 = 0$ for identifiability. Similar to logistic regressions, non-collapsibility issue also arises in the BT model, and $\bphi$ and $\bpsi$ are different in general. We will consider both estimands in the following, but we note that when the conditional BT model is indeed correct, it is natural to consider $\bphi$ especially with covariate shifts.

\paragraph{Identification.} We consider the situation without covariate shift first and write our estimands as functionals of the observed data distribution $P$ under appropriate assumptions. We use $Y^{kl}$ to denote the counterfactual outcome of whether the player with smaller index wins had pair $(k,l)$ been drawn. Under the following Assumption~\ref{cond: no covariate shift}, $\bphi$ and $\bpsi$ can be identified from $P$ as both $Q_{\bX}$ and $q_{kl}$ can be identified.

\begin{assumption}\label{cond: no covariate shift}
\sloppy Under SUTVA and suppose that (a) identifiability: $P_{\bX} = Q_{\bX}$, and $\EE_P[Y^{kl}|X=\bx] = q_{kl}(\bx)$; (b) no unmeasured confounding: $Y^{kl} \perp A |X$, for all $(k,l) \in \mathcal{A}$; and (c) positivity: $P(A=a|\bX=\bx) > \delta$ for some $\delta >0$ and all $a \in \mathcal{A}$ and $\bx \in \mathcal{X}$.
\end{assumption}
We allow the conditional BT model to be misspecified, and assume a locally nonparametric model for the target distribution $Q_{Y|(A,\bX)}$ and $Q_{\bX}$. Consequently, we will assume that the observed data distribution $P \in \mathcal{M}$, where $\mathcal{M}$ is a locally nonparametric model of distributions of $(\bX,A,Y)$. With a slight abuse of notation, we define our parameter of interest as a functional of the observed data distribution, $\bPhi: \mathcal{M} \rightarrow \mathbb{R}^{K-1}$, such that $\bPhi(P) = \EE_{P_{\bX}}[\btheta^*(\bX)]$ and $\btheta^*(\bx)$ satisfies \eqref{eq: EE given x} with $q_{kl}(\bx)$ replaced by $\EE_P[Y|A=(k,l),\bX = \bx]$. Under Assumption~\ref{cond: no covariate shift}, $\bPhi(P)$ also coincides with $\EE_{Q_{\bX}}[\btheta^*(\bX)]$. We similarly define a parameter $\bPsi: \mathcal{M} \rightarrow \mathbb{R}^{K-1}$ for the alternative estimand. Specifically, $\bPsi(P)$ solves \eqref{eq: marginal EE} with $q_{kl}$ replaced by $\EE_P[\EE_P[Y|A=(k,l),\bX]]$.

Oftentimes, one may wish to evaluate the players under specific contexts, for example, accuracy or toxicity of LLMs when responding to certain types of prompts in a specific task.  In this situation with covariate shift where $Q_{\bX}$ and $P_{\bX}$ differ, we assume that we have access to additional unlabeled samples of $\bX$ drawn from $Q_{\bX}$. Defining an indicator $S$ taking value 1 for labeled samples drawn from $P$ and 0 for unlabeled samples drawn from $Q_{\bX}$, a generic data unit now takes the form $\bO = (S,\bX,SA,SY) \sim \PP$ under which $(\bX,A,Y)|S=1 \sim P$ and $\bX|S=0 \sim Q_{\bX}$. To ensure identifiability under covariate shift, we replace condition (a) in Assumption~\ref{cond: no covariate shift} with the following:
\begin{assumption}\label{cond: with covariate shift}
   $Q_{\bX}$ is absolutely continuous with respect to $P_{\bX}$, and there exists constant $c>0$ such that $c^{-1} < dQ_{\bX}/dP_{\bX}(\bx) < c$ for $P_{\bX}$-a.e. $\bx$. Moreover, $\EE_P[Y^{kl}|X=\bx] = q_{kl}(\bx)$.
\end{assumption}
Without imposing model assumptions, we assume $\PP \in \mathcal{M}_f$, where $\mathcal{M}_f$ denotes a locally nonparametric model of distributions of $\bO$. We define a functional $\bPhi_f: \mathcal{M}_f \rightarrow \mathbb{R}^{K-1}$ such that $\bPhi_f(\PP) = \EE_\PP[\btheta^*(X)|S=0]$ where $\btheta^*(\bx)$ satisfies \eqref{eq: EE given x} with $q_{kl}(\bx) = \EE_\PP[Y|A=(k,l),\bX=\bx, S=1]$ under Assumptions~\ref{cond: no covariate shift}(b) and \ref{cond: with covariate shift}. We also define $\bPsi_f: \mathcal{M}_f \rightarrow \mathbb{R}^{K-1}$ such that $\bPsi(\PP)$ satisfies \eqref{eq: marginal EE} with $q_{kl} = \EE_\PP[\EE_\PP[Y|A=(k,l),\bX, S=1] | S = 0]$ under the same assumptions. In Appendix~\ref{app:known density ratio}, we consider an alternative identification strategy where we assume that the density ratio $dQ_{\bX}/dP_{\bX}$ is known up to a normalizing factor.

\section{One-step estimation based on the efficient influence function}\label{sec: nonparam estimation}
We first consider the setting without covariate shift where we observe an i.i.d. sample of size $n$ drawn from $P$, $\{(\bX_i,A_i,Y_i)\}_{i=1}^n$. We derive the efficient influence function (EIF) of $\bphi$ and $\bpsi$ under a locally nonparametric model of $P$. Let $m_{kl}(\bx) := \EE_P[Y|A=(k,l),\bX=\bx]$ denote the conditional mean of $Y$ and define $m_{lk}(\bx) = 1 - m_{kl}(\bx)$, for $1 \leq k < l \leq K$. We further define a vector $\bm(\bx) \in \mathbb{R}^{(K-1)^2}$ such that $\bm(\bx) = (\bm_2(\bx)^\top,\ldots,\bm_K(\bx)^\top)^\top$ and $\bm_k(\bx)^\top = (m_{k1}(\bx),\ldots,m_{k(k-1)}(\bx),m_{k(k+1)}(\bx),\ldots,m_{kK}(\bx))$. The vector $\bm(\bx)$ simply collects the conditional winning probabilities in all pairwise comparisons given $\bX = \bx$. Let $\bm = \EE_P[\bm(\bX)]$. We also define the propensity score as $\pi(a|\bx) := P_{A|\bX}(A=a|\bX=\bx)$ for $a \in \calA$.

To simplify the presentation, we introduce the following notations. First we use $\bU \in \mathbb{R}^{K-1}$ to denote the estimating equations in the form of \eqref{eq: EE given x} and \eqref{eq: marginal EE} and use $\Lambda \in \mathbb{R}^{(K-1)\times (K-1)^2}$ to denote a matrix involving its partial derivatives. Let $\bm^\dagger \in \mathbb{R}^{(K-1)^2}$ denote a generic vector in the form of $\bm$ and $\bm(\bx)$ and $\btheta^\dagger \in \mathbb{R}^{K-1}$ denote a generic vector of strengths. For $2 \leq k \leq K$, the $(k-1)$-th entry of $\bU$ is defined as
\begin{equation*}
    \bU_{k-1}(\btheta^\dagger,\bm^\dagger; \rho) = \sum_{l \neq k} \rho_{kl}\left\{\sigma(\theta^\dagger_k - \theta^\dagger_l) - m^\dagger_{kl}\right\}.
\end{equation*}
With this definition, we can now rewrite \eqref{eq: EE given x} as $\bU(\btheta^*(\bx),\bm(\bx);\rho(\bx)) = \boldsymbol{0}$ for a.e. $\bx$ under the identifiability assumptions, and similarly rewrite \eqref{eq: marginal EE} as $\bU(\bpsi,\bm;\rho) = \boldsymbol{0}$. The matrix $\Lambda$ involving the partial derivatives of $\bU$ is defined as
\begin{equation*}
    \Lambda(\btheta^\dagger,\bm^\dagger;\rho) = \left\{\frac{\partial \bU(\btheta^\dagger,\bm^\dagger;\rho)}{\partial \btheta^\dagger}\right\}^{-1}
    \left\{\frac{\partial \bU(\btheta^\dagger,\bm^\dagger;\rho)}{\partial \bm^\dagger}\right\}.
\end{equation*}
For brevity, we present the explicit expressions for the partial derivatives $\partial \bU/\partial \btheta^\dagger$ and $\partial \bU/\partial \bm^\dagger$ in Appendix~\ref{app:proof}. We will often suppress the dependence on $\rho$ when it is clear from the contexts which given $\rho$ we are considering.
\begin{proposition}[EIF of $\bPhi$ and $\bPsi$]\label{prop: EIF no shift}
    Under Assumption~\ref{cond: no covariate shift}, the EIF of $\bPhi: \mathcal{M} \rightarrow \mathbb{R}^{K-1}$ is
    \begin{equation*}
    D_{\bphi}(\bx,a,y) = - \Lambda(\btheta^*(\bx),\bm(\bx);\rho(\bx))
    \tau(\bx,a,y) + \btheta^*(\bx) - \bphi,
\end{equation*}
where
\begin{align}
    \tau(\bx,a,y) &= (\tau_2(\bx,a,y)^\top, \ldots, \tau_{K}(\bx,a,y)^\top)^\top; \nonumber \\
    \tau_k(\bx,a,y) &= (\tau_{k1}(\bx,a,y),\ldots,\tau_{k(k-1)}(\bx,a,y),\tau_{k(k+1)}(\bx,a,y),\ldots,\tau_{kK}(\bx,a,y))^\top; \nonumber \\
    \tau_{kl}(\bx,a,y) &=  (-1)^{I\{k < l\}+1} \frac{I\left\{a = (k\wedge l, k\vee l)\right\}}{\pi((k\wedge l, k\vee l)|\bx)} \left\{y - m_{(k\wedge l)(k\vee l)}(\bx)\right\}. \label{eq: definition of tau}
\end{align}
The EIF of $\bPsi: \mathcal{M} \rightarrow \mathbb{R}^{K-1}$ is
\begin{equation*}
    D_{\bpsi}(\bx,a,y) = - \Lambda(\bpsi,\bm; \rho)\left\{\tau(\bx,a,y) + \bm(\bx) - \bm\right\}.
\end{equation*}
\end{proposition}

Given initial estimates $\widehat{m}_{kl}(\bx)$ of the conditional mean functions $m_{kl}(\bx)$, which coincides with $q_{kl}(\bx)$ under Assumption~\ref{cond: no covariate shift}, one can obtain estimates $\widehat{\btheta}(\bx)$ by solving the system of equations in \eqref{eq: EE given x} pointwise in $\bx$ and construct a plug-in estimator by marginalizing $\widehat{\btheta}(\bx)$ over the empirical distribution of $\bX$. However, this simple plug-in estimator may contain non-negligible bias for the purpose of inference. Instead, we will construct a one-step estimator based on the EIF. In addition to the estimated conditional mean functions $\widehat{\bm}(\bx)$ and the resulting $\widehat\btheta(\bx)$ estimate, we also need estimates of $\pi(a|\bx) $, which we denote as $\widehat\pi(a|\bx)$, for $a \in \calA$. Define $\widehat\tau$ in the same way as $\tau$ but with all nuisance functions replaced by their corresponding estimates. Define the following estimator $\widehat\bphi$
\begin{equation*}
    \widehat\bphi = n^{-1} \sum_{i=1}^n \left\{ \widehat\btheta(\bX_i) - \widehat\Lambda(\widehat\btheta(X_i),\widehat\bm(X_i))
    \widehat\tau(\bX_i,A_i,Y_i) \right\}.
\end{equation*}
A similar one-step estimator can be constructed for $\bpsi$, and we present the details in Appendix~\ref{app:psi}.

Next, we consider situations with covariate shift. In addition to the i.i.d. sample from $P$, we observe another unlabeled i.i.d. sample $\{\bX_i\}_{i=n+1}^{n + m}$ drawn directly from $Q_{\bX}$. More compactly, we have i.i.d. copies of $(S_i,\bX_i,S_iA_i,S_iY_i)$ for $1\leq i \leq N$ drawn from $\PP$, where $N = n + m$. Under Assumptions~\ref{cond: no covariate shift}(b)-(c) and \ref{cond: with covariate shift}, $\bphi$ and $\bpsi$ can be identified from the distribution of both the labeled and unlabeled data. We refer to this scenario as one with \emph{data fusion}. 

\begin{proposition}[EIF of $\bPhi_f$ and $\bPsi_f$ with data fusion]\label{prop: EIF with shift}
The EIFs of $\bPhi_f: \mathcal{M}_f \rightarrow \mathbb{R}^{K-1}$ and $\bPsi_f: \mathcal{M}_f \rightarrow \mathbb{R}^{K-1}$ under data fusion are:
\begin{align*}
    D_{\bphi,f}(s,\bx,a,y) &= - \frac{s}{Pr(S=1)}\frac{dQ_{\bX}}{dP_{\bX}}(\bx)\Lambda(\btheta^*(\bx),\bm(\bx);\rho(\bx)) \tau(\bx,a,y) + \frac{(1-s)}{Pr(S=0)}\left\{\btheta^*(\bx) - \bphi\right\}; \\
    D_{\bpsi,f}(s, \bx,a,y) &= - \Lambda(\bpsi,\bm;\rho) \left\{\frac{s}{Pr(S=1)}\frac{dQ_{\bX}}{dP_{\bX}}(\bx)\tau(\bx,a,y) + \frac{(1-s)}{Pr(S=0)}\left(\bm(\bx) - \bm\right)\right\}.
\end{align*}
\normalsize
\end{proposition}
Let $\widehat{w}$ denote an estimate of the unknown density ratio $dQ_{\bX}/dP_{\bX}$, and let $\bar{S}$ be the sample mean of $S_i$, $\sum_{i=1}^{N}S_i/N$. With other nuisance function estimates $\widehat\btheta$, $\widehat\bm$ and $\widehat\pi$, we can construct the following one-step estimator of $\bphi$ under covariate shift in the data fusion setting:
\begin{align*}
    \widehat\bphi_f &= N^{-1}\sum_{i=1}^{N} - \frac{S_i}{\bar S}\widehat{w}(\bX_i)\widehat\Lambda(\widehat\btheta(\bX_i),\widehat\bm(\bX_i))
    \widehat\tau(\bX_i,A_i,Y_i)  + N^{-1}\sum_{i=1}^{N}\frac{(1-S_i)}{1-\bar S}\widehat\btheta(\bX_i),
\end{align*}
\normalsize
where $\widehat\tau$ is again defined in the same way as $\tau$ but with all unknown nuisance functions replaced with their estimates. Given a plug-in estimator $\widetilde{\bpsi}$ of $\bpsi$, we can similarly define a one-step estimator of $\bpsi$, $\widehat\bpsi = \widetilde{\bpsi} + N^{-1}\sum_{i=1}^{N} D_{\bpsi,f}(S_i,\bX_i,S_iA_i,S_iY_i)$. The details are deferred to Appendix~\ref{app:psi}.

It is worth pointing out that even when the density ratio $dQ_{\bX}/dP_{\bX}$ is not lower bounded away from 0, the EIFs in Proposition~\ref{prop: EIF with shift} remain as valid influence functions, although they may no longer be efficient. We refer the readers to \cite{li2023efficient} for more discussion on this issue.

\section{Inference under correctly specified conditional BT model}\label{sec: cond BT estimation}
In the previous sections, we use the conditional BT model as a potentially misspecified working model, and derive EIF-based estimators of $\bphi$ and $\bpsi$. Notably, we need the positivity assumption to hold for all $a \in \calA$, and the EIF involves $\pi(a|\bx)$ for all $a \in \calA$. Implementing the one-step estimators requires estimation of the $K(K-1)/2$-dimensional conditional multinomial probability vector as a function of $\bx$. While this is necessary as $\bphi$ and $\bpsi$ depend on the winning probabilities in all pairwise comparisons, with large $K$, nonparametric estimation of $\pi$ can be difficult. Moreover, $\pi(a|\bx)$ may be small for a given $a$, leading to numerical instability. One mitigation is to assume that the two players in a comparison are sampled independently by design in the observed data, in which case only a $K$-dimensional multinomial probability vector needs to be estimated. In this section, we explore another scenario where the conditional BT model is correctly specified. We show that in this case, $\btheta(\bx)$ can be identified with as few as $(K-1)$ pairwise comparisons, and consequently the one-step estimators involve fewer nuisance functions to be estimated.

\begin{remark}
The motivation behind specifying a conditional model is as follows. When the winning probabilities in the pairwise comparisons depend on the covariates, even if the data generating distribution belongs to the marginal BT model for one covariate distribution, it may no longer belong to the marginal BT model under a different covariate distribution. In such situations with potential covariate shift, it is natural to specify a conditional model instead. Furthermore, as the BT model is not collapsible in general, even if the conditional distribution of all pairwise comparisons given contextual covariates belongs to the conditional BT model, the marginal distribution may not belong to the marginal BT model.
\end{remark}

Consider the situation without covariate shift such that $Q \in \mathcal{P}$ and $P_{\btheta^*} = Q$. We see that $\btheta^*(\bx)$ satisfies a system of linear equations involving $\bm(\bx)$ --- under correct model specification and identifiability assumptions $\theta_k^*(\bx) - \theta_l^*(\bx) = \sigma^{-1}(m_{kl}(\bx))$ where $\sigma^{-1}(\cdot)$ is the logit function. To make the presentation concise, we define a comparison matrix $\Gamma \in \mathbb{R}^{J \times (K-1)}$ that encodes a subset of comparisons among the $K$ players, where each row corresponds exactly to one pairwise comparison. The number of comparisons, $J$, can be as small as $K-1$ to ensure the identification of $\btheta^*$. Specifically, if players $k_j$ and $l_j$ are compared in match $j$ with $k_j < l_j$, we set $\Gamma_{j,k_j-1} = 1$ and $\Gamma_{j,l_j-1} = -1$. As a result, each row of $\Gamma$ contains exactly one $1$, one $-1$ and zero elsewhere. For comparisons involving player 1, the corresponding row in $\Gamma$ contains only one nonzero entry, $-1$, and we set $k_j=1$. For example, when $K=4$, the row $(1,0,-1)$ indicates a comparison between players 2 and 4, whereas a row $(0,-1,0)$ indicates a comparison between players 1 and 3. Subsequently, the linear system can be written compactly in a matrix form $\Gamma\btheta(\bx) = \sigma^{-1}(\bm_\Gamma(\bx))$, where $\bm_\Gamma = (m_{k_1l_1}(\bx),\ldots,m_{k_Jl_J}(\bx))$ and $\sigma^{-1}$ is applied element-wise. When $\Gamma$ is invertible or more generally has full column rank, we have $\btheta^*(\bx) = \Gamma^{-1}\sigma^{-1}(\bm_\Gamma(\bx))$ with $\Gamma^{-1}$ denoting the (pseudo)inverse, and therefore $$\bphi = \EE_Q[\Gamma^{-1}\sigma^{-1}(\bm_\Gamma(\bX))].$$ In Appendix~\ref{app:graph}, we show that the matrix $\Gamma$ is related to the incidence matrix of a comparison graph and that full column rank of $\Gamma$ corresponds to the connectedness of this graph. 

\begin{remark}
In general, when more than $K-1$ pairs are observed under $P$, the choice of $\Gamma$ is not unique. This leads to multiple ways for identification. In fact, for identification purposes, $\Gamma$ can be a function of $\bx$ as long as $\Gamma(\bx)$ is invertible for all $\bx$. This means that at different covariate values, the pairs used to identify $\btheta(\bx)$ can be different theoretically. However, as we will see, the one-step estimator will involve estimation of $\pi((k_j,l_j)|\bx)$. This is generally infeasible with continuous covariates without further assumptions if $\Gamma$ is not ``smooth" in $\bx$. Therefore, we have assumed a fixed $\Gamma$ independent of $\bx$.
\end{remark}

Let $\calA_J := \{(k_j,l_j): 1\leq j \leq J\}$. We impose the following assumptions.
\begin{assumption}\label{cond: no shift cond}
\sloppy Under SUTVA and suppose that (a) identifiability: $P_{\bX} = Q_{\bX}$, and $\EE_P[Y^{kl}|X=\bx] = q_{kl}(\bx)$ for $(k,l) \in \calA_J$; (b) no unmeasured confounding: $Y^{kl} \perp A |X$, for all $(k,l) \in \mathcal{A}_J$; and (c) positivity: $P(A=a|\bX=\bx) > \delta$ for some $\delta >0$ and all $a \in \mathcal{A}_J$ and $\bx \in \mathcal{X}$.
\end{assumption}

\begin{assumption}\label{cond: with shift cond}
$Q_{\bX}$ is absolutely continuous with respect to $P_{\bX}$, and there exists constant $c>0$ such that $c^{-1} < dQ_{\bX}/dP_{\bX}(\bx) < c$ for $P_{\bX}$-a.e. $\bx$. Moreover, $\EE_P[Y^{kl}|X=\bx] = q_{kl}(\bx)$ for $(k,l) \in \calA_J$.
\end{assumption}

\begin{assumption}\label{cond: gamma full rank}
$\Gamma \in \mathbb{R}^{J \times (K-1)}$ has full column rank.
\end{assumption}

\noindent We note that, using this identification strategy under correct model specification, Assumption~\ref{cond: no shift cond}(c) only requires positivity for pairs $a \in \calA_J$. Thus, we do not need all pairwise comparisons to be observed or annotated. Furthermore, $J$ can be as small as $K-1$ for Assumption~\ref{cond: gamma full rank} to hold. The following result gives an influence function of $\bPhi$ without covariate shift and $\bPhi_f$ with covariate shift.
\begin{proposition}[IF of $\bPhi$ and $\bPhi_f$ under correctly specified conditional BT model]\label{prop: IF conditional BT}
    \sloppy Suppose that $Q \in \mathcal{P}$. Under Assumptions~\ref{cond: no shift cond} and \ref{cond: gamma full rank}, the following function is an influence function of $\bPhi: \calP \rightarrow \mathbb{R}^{K-1}$
\begin{equation*}
    D_{cond}: (\bx,a,y) \mapsto \Gamma^{-1}\tilde\tau(\bx,a,y) + \Gamma^{-1}\sigma^{-1}(\bm_\Gamma(\bx)) - \bphi,
\end{equation*}
where $\tilde\tau(\bx,a,y) = (\tilde\tau_1(\bx,a,y),\ldots, \tilde\tau_J(\bx,a,y)) \in \mathbb{R}^J$ and, with $\tau$ defined in \eqref{eq: definition of tau}, 
$$\tilde\tau_j(\bx,a,y) = \tau_{k_jl_j}(\bx,a,y)m_{k_jl_j}(\bx)^{-1}(1-m_{k_jl_j}(\bx))^{-1}.$$
Under Assumptions~\ref{cond: no shift cond}(b)-(c), \ref{cond: with shift cond} and \ref{cond: gamma full rank} in the data fusion setting, the following function is an influence function of $\bPhi_f$:
\begin{equation*}
    D_{cond,f}: (s,\bx,a,y) \mapsto \frac{dQ_{\bX}}{dP_{\bX}}(\bx)\frac{s\Gamma^{-1}\tilde\tau(\bx,a,y)}{Pr(S=1)} + \frac{1-s}{Pr(S=0)}\left\{\Gamma^{-1}\sigma^{-1}(\bm_\Gamma(\bx)) - \bphi\right\}.
\end{equation*}
\end{proposition}
Given estimates of the nuisance functions, we can construct the following plug-in estimators and corresponding one-step estimators
\begin{align*}
    \widetilde\bphi_{cond} &= n^{-1}\sum_{i=1}^n \Gamma^{-1}\sigma^{-1}(\widehat\bm_\Gamma(\bX_i)), \quad \widehat\bphi_{cond} = \widetilde\bphi_{cond} + n^{-1}\sum_{i=1}^n \widehat D_{cond} (\bX_i,A_i,Y_i);\\
    \widetilde\bphi_{cond,f} &= m^{-1}\sum_{i=1}^N S_i\Gamma^{-1}\sigma^{-1}(\widehat\bm_\Gamma(\bX_i)), \quad \widehat\bphi_{cond,f} = \widetilde\bphi_{cond,f} + N^{-1}\sum_{i=1}^N \widehat D_{cond,f} (S_i,\bX_i,S_iA_i,S_iY_i).
\end{align*}
Although we focus on $\bphi$ here, we present analogous results for $\bpsi$ under correctly specified conditional BT model in Appendix~\ref{app:psi} and derive corresponding estimators.

\subsection{Improved efficiency by leveraging more pairs}

We have shown that if the conditional BT model is indeed correctly specified, $\bphi$ can be identified and estimated using as few as $K-1$ pairwise comparisons. If more pairwise comparisons are observed under $P$, however, leveraging all observed pairs will improve the estimation efficiency (although this will also necessitate estimating more propensity scores.) To this end, we now derive the efficient influence functions of $\bPhi$ and $\bPhi_f$ under the conditional BT model by projecting the influence functions in Proposition~\ref{prop: IF conditional BT} onto the tangent space. We note that under the conditional BT model, the conditional distributions of $Y|A=a,\bX$ for different values of $a$ are no longer variational independent, as they are parametrized by the lower-dimensional vector $\btheta$. We present the form of the tangent space in Appendix~\ref{app:graph}

Again, to make the presentation concise, we define a full comparison matrix $\Gamma_* \in \mathbb{R}^{\frac{(K-1)K}{2} \times (K-1)}$ in the same way as $\Gamma$, except that $\Gamma_*$ now encodes all possible pairwise comparisons among $K$ players. That is, for any $a = (k,l) \in \calA$, there exists $1 \leq j \leq K(K-1)/2$ such that $(\Gamma_*)_{j,k-1} = 1$ and $(\Gamma_*)_{j,l-1} = -1$. In other words, we can now order all pairs according to the rows of $\Gamma_*$, $\calA = \{(k_{*j},l_{*j}): 1\leq j \leq K(K-1)/2\}$. Define a weight matrix $W_*(\bx)$ that is a $K(K-1)/2$ by $K(K-1)/2$ diagonal matrix, with the $j$-th diagonal element equal to $m_{k_{*j}l_{*j}}(\bx)(1-m_{k_{*j}l_{*j}}(\bx))\pi((k_{*j},l_{*j})|\bx)$ and zeros off-diagonal. Note that for unobserved pairs, the weight is automatically set to 0 as $\pi((k_{*j},l_{*j})|\bx)$ is 0. 

\begin{proposition}[EIF of $\bphi$ under conditional BT model]\label{prop: EIF cond}
\sloppy Suppose that $Q \in \mathcal{P}$. Under Assumptions~\ref{cond: no shift cond} and \ref{cond: gamma full rank}, the EIF of $\bPhi : \calP \rightarrow \mathbb{R}^{K-1}$ at $P$ is
\begin{equation*}
    D^*_{cond}: (\bx,a,y) \mapsto \left(\Gamma_*^\top W_*(\bx) \Gamma_*\right)^{-1}\Gamma_*^\top v(\bx,a,y) + \btheta^*(\bx) - \bphi,
\end{equation*}
where $v(\bx,a,y) \in \mathbb{R}^{K(K-1)/2}$ with the $j$-th entry equal to $I\{a=(k_{*j},l_{*j})\}I\{\pi((k_{*j},l_{*j})|\bx) > 0\}\{y-m_{k_{*j}l_{*j}}(\bx)\}$. Under Assumptions~\ref{cond: no shift cond}(b)-(c), \ref{cond: with shift cond} and \ref{cond: gamma full rank} in the data fusion setting, the EIF of $\bPhi_f$ at $\PP$ is
\begin{equation*}
    D^*_{cond,f}: (s,\bx,a,y) \mapsto \frac{S}{Pr(S=1)} \frac{dQ_{\bX}}{dP_{\bX}}(\bx)\left(\Gamma_*^\top W_*(\bx) \Gamma_*\right)^{-1}\Gamma_*^\top v(\bx,a,y) + \frac{1-S}{Pr(S=0)}\left\{\btheta^*(\bx) - \bphi\right\}.
\end{equation*}
\end{proposition}

\begin{corollary}\label{corollary: minimal pairs}
Suppose that the assumptions in Proposition~\ref{prop: EIF cond} hold. Further suppose that for some $\Gamma \in \mathbb{R}^{J \times (K-1)}$ with $J = K-1$, $P(A = a|\bX = \bx) = 0$ for all $\bx \in \mathcal{X}$ for all $a \notin \calA_J$. Then $D_{cond} = D_{cond}^*$ and $D_{cond,f} = D_{cond,f}^*$.
\end{corollary}
\noindent The EIFs no longer depend on the potentially non-unique choice of $\Gamma$, but they now involve all (non-zero) propensity scores. Moreover, if there are only $K-1$ pairs observed under $P$ and $\Gamma$ has full column rank, the EIFs coincide with the influence functions in Proposition~\ref{prop: IF conditional BT}. This is intuitive as $\Gamma$ already encodes all the information on $\btheta$ contained in the data.

Implementing the one-step estimators based on the EIFs requires estimation of all non-zero propensity scores. In estimating the other nuisance functions, we also need to ensure that they indeed correspond to some distribution $\hat{P}$ in the conditional BT model. One way to achieve this is to first estimate $m_{kl}(\bx)$ nonparametrically and obtain some initial estimates $\widehat{m}_{kl}^{init}(\bx)$, and then obtain $\widehat{\btheta}(\bx)$ via \eqref{eq: EE given x}. In forming the estimators, we estimate the condition mean via $\widehat{m}_{kl}(\bx) = \sigma(\widehat\theta_k(\bx) - \widehat\theta_l(\bx))$. The explicit forms of the estimators are as follows
\begin{align*}
    \widehat\bphi_{cond}^* &= n^{-1}\sum_{i=1}^n \left\{\left(\Gamma_*^\top \widehat W_*(\bX_i) \Gamma_*\right)^{-1}\Gamma_*^\top \widehat v(\bX_i,A_i,Y_i) + \widehat\btheta(\bX_i) \right\}; \\
    \widehat\bphi_{cond,f}^* &= N^{-1}\sum_{i=1}^{N} \frac{S_i}{\bar S}\widehat{w}(\bX_i) \left(\Gamma_*^\top \widehat W_*(\bX_i) \Gamma_*\right)^{-1}\Gamma_*^\top \widehat v(\bX_i,A_i,Y_i)  + N^{-1}\sum_{i=1}^{N}\frac{(1-S_i)}{1-\bar S}\widehat\btheta(\bX_i).
\end{align*}

\section{Theoretical guarantees}
\label{sec:theoretical guarantees}
We present the theoretical properties of the proposed estimators $\widehat{\bphi}_f$, $\widehat{\bpsi}_f$ and $\widehat{\bphi}^*_{cond,f}$ under covariate shift in this section, and defer readers' attention to the appendix for properties of other estimators. Let $\lVert \cdot \rVert$ denote the $L_0^2(\PP)$ norm and $P_{N}$ denote the empirical measure. In addition, we define the $L_0^2(\PP)$ norm of a random matrix $\Lambda$ using the operator norm that $\lVert \Lambda(\bX)\rVert = \sqrt{E_{\PP}\left[\lVert \Lambda(\bX)\rVert^2_{F} \right]}$. 

\begin{theorem}[\textit{Efficient 
estimation of $\bphi$ under covariate shifts}] \label{thm: dr_phi_shift}
    \sloppy Suppose the conditional BT model is misspecified, and nuisance functions  $\widehat{\pi}$, $\widehat{\bm}$, and $\widehat{w}$ were estimated via cross-fitting. Under Assumptions~\ref{cond: no covariate shift}(b)-(c) and \ref{cond: with covariate shift}, we have
    \begin{align*}
        \widehat{\bphi}_f - \bphi &= P_{N}D_{\bphi,f} + O_{{\PP}}\big( \lVert \hat{w}(\bx) - w(\bx) \rVert \lVert \widehat{\bm}(\bx) - \bm(\bx) \rVert  + \lVert \widehat{\bm}(\bx) - \bm(\bx) \rVert^2\\
    & \quad  +  \lVert \widehat{\pi}(\bx) - \pi(\bx)\rVert \lVert \widehat{\bm}(\bx) - \bm(\bx)\rVert + \lVert \widehat{\Lambda}(\bx) - \Lambda(\bx)\rVert  \lVert \widehat{\bm}(\bx) - \bm(\bx)\rVert  \big).
    \end{align*} 
    Moreover, if the nuisance functions were estimated such that the sum of the four product terms above is $o_{\PP}(1/\sqrt{N})$, then the proposed estimator $\widehat{\bphi}_f$ is consistent, asymptotically normal and achieve the semiparametric efficiency bound. That is,
    $$\sqrt{N}(\widehat{\bphi}_f - \bphi) \rightarrow_d \mathrm{N}(\mathbf{0}, \mathrm{cov}_{{\PP}}(D_{\bphi,f})).$$
\end{theorem}

\begin{remark}
    The use of cross-fitting \citep{zheng2010asymptotic} in estimating nuisance functions ensures that appropriate empirical process and consistency conditions are both met (Lemma 19.24 of \citep{van2000asymptotic}). Alternatively, one may assume that the estimated nuisance functions belong to a fixed Donsker class with probability tending to 1.
\end{remark}

Theorem~\ref{thm: dr_phi_shift} suggests that $\widehat{\bphi}_f$ is consistent if $\bm$ can be estimated consistently. Moreover, $\widehat{\bphi}_f$ achieves asymptotic linearity and efficiency when the product of the convergence rates of nuisance functions are root-n rate. One way to achieve efficiency is to estimate $\bm$ using a correctly specified model, which in turn, would only require other nuisance functions to be consistently estimated at arbitrary slow rates. Alternatively, efficiency can also be obtained if all nuisance functions were estimated at $n^{-1/4}$ rate. Various data-adaptive methods can attain this required rate, such as generalized additive model \citep{hastie2017generalized} and methods of sieves \citep{shen1997methods}. Moreover, all of the proposed estimators introduced in Section~\ref{sec: nonparam estimation} enjoy asymptotic normality and achieve the corresponding semiparametric efficiency bound under similar conditions. To illustrate, we present the result for estimating $\bpsi$ under covariate shift below.

\begin{theorem}[\textit{Efficient and doubly robust 
estimation of $\bpsi$ under covariate shifts} ] \label{thm: dr_psi_shift}
    Suppose the conditional BT model is misspecified, and nuisance functions $\hat{w}$, $\widehat{\pi}$ and $\widehat{\bm}$ were estimated via cross-fitting. Under Assumptions~\ref{cond: no covariate shift}(b)-(c) and~\ref{cond: with covariate shift} , we have
    \begin{align*}
        \widehat{\bpsi}_f - \bpsi &= P_{N}D_{\bpsi,_f} + O_{{\PP}}( \lVert \hat{w}(\bx) - w(\bx) \rVert \lVert \widehat{\bm}(\bx) - \bm(\bx) \rVert  +  \lVert \widehat{\pi}(\bx) - \pi(\bx)\rVert \lVert \widehat{\bm}(\bx) - \bm(\bx)\rVert  ).
    \end{align*} 
    Moreover, if the nuisance functions were estimated such that the sum of the two product terms above is $o_{\PP}(1/\sqrt{N})$, then the proposed estimator $\widehat{\bpsi}_f$ is consistent, asymptotically normal and achieve the semiparametric efficiency bound. That is,
    $$\sqrt{N}(\widehat{\bpsi}_f - \bpsi) \rightarrow_d \mathrm{N}(\mathbf{0}, \mathrm{cov}_{{\PP}}(D_{\bpsi,_f})).$$
\end{theorem}
Comparing to results in Theorem~\ref{thm: dr_phi_shift}, the estimation of $\psi$ enjoys an additional appealing feature of being doubly robust. That is, the estimator $\widehat{\bpsi}_f$ will be consistent if either $\bm$ is consistently estimated, or $w$ and $\pi$ are consistently estimated. We provide the results for other estimators introduced in Section~\ref{sec: nonparam estimation} in  Appendix~\ref{app:other estimator}. Next, we present the result for estimating $\bphi$ under covariate shifts when the conditional BT model holds.

\begin{theorem}[\textit{Efficient estimation of $\bphi$ under covariate shifts and conditional BT} ] \label{thm: dr_phi_shift_cond}
   Suppose the conditional BT model is correctly specified, and nuisance functions $\hat{w}$, $\widehat{\pi}$ and $\widehat{\bm}$ were estimated via cross-fitting. Under Assumptions~\ref{cond: no shift cond}(b)-(c), \ref{cond: with shift cond} and \ref{cond: gamma full rank}, we have
    \begin{align*}
        \widehat{\bphi}^*_{cond,f} - \bphi &= P_{N}D^*_{cond,f} + O_{{\PP}}( \lVert \widehat{\pi}(\bx) - \pi(\bx)\rVert \lVert \widehat{\bm}(\bx) - \bm(\bx)\rVert \\
        & \quad +\lVert \widehat{w}(\bx) - w(\bx)\rVert \lVert \widehat{\bm}(\bx) - \bm(\bx)\rVert + \lVert \widehat{\bm}(\bx) - \bm(\bx)\rVert^2  ).
    \end{align*} 
    Moreover, if the nuisance functions were estimated such that the sum of the product terms above is $o_{\PP}(1/\sqrt{N})$, then the proposed estimator $\widehat{\bphi}^*_{cond,f}$ is consistent and asymptotically normal. That is,
    $$\sqrt{N}(\widehat{\bphi}^*_{cond,f}- \bphi) \rightarrow_d \mathrm{N}(\mathbf{0}, \mathrm{cov}_{{\PP}}(D^*_{cond,f})).$$
\end{theorem}
When only $K-1$ comparisons are observed, the conditions above apply only to $\bm_{\Gamma}$. Since we estimate the condition mean via $\widehat{m}_{kl}(\bx) = \sigma(\widehat\theta_k(\bx) - \widehat\theta_l(\bx))$, the terms above also are equivalent to the ones where we replace $\lVert \widehat{\bm}(\bx) - \bm(\bx)\rVert $ by $\lVert \widehat{\btheta}(\bx) - \btheta(\bx)\rVert $. We provide additional results and corresponding proofs in  Appendix~\ref{app:other estimator}. 

\section{Numerical experiments}\label{sec: experiments}

\subsection{Simulation studies}\label{subsec: sim}

We evaluate the performance of our proposed inferential procedures in simulation studies. We first consider a setting (Setting I) where the underlying data generating distribution does not belong to the marginal nor the conditional BT model and there is covariate shift. The number of players $K$ is 3, and the covariate vector $\bX \in \mathbb{R}^2$. In the labeled data, $X_1 \sim N(0,0.5^2)$, $X_2 \sim \textnormal{Bernoulli(0.5)}$, and $X_1$ is independent of $X_2$. We set $\pi(a|\bx) = 1/3$ for $a \in \{(1,2), (1,3), (2,3)\}$ and all $\bx$. The winning probabilities in the pairwise comparisons are set as $m_{12}(\bx) = 0.5 + 0.2\sin\{1.5(x_1 + x_2)\}$, $m_{13}(\bx) = \sigma(0.3x_1(x_2-1))$, and $m_{23}(\bx) = \sigma(0.2x_1^2-0.5)$. Given $\bX$ and $A$, the outcome $Y$ is generated according to $\textnormal{Bernoulli}(m_A(\bX))$. In the unlabeled target data, we generate $X_1$ and $X_2$ independently with $X_1 \sim \textnormal{Uniform}(0,0.5)$ and $X_2 \sim \textnormal{Bernoulli}(0.4)$. In defining $\bphi$ and $\bpsi$, we set $\rho_{kl} = 1/3$ for all $k\neq l$, which represents a fair comparison where the players have an equal chance to face each other under different contexts. This results in the true values $(\bphi_1,\bphi_2,\bphi_3) = (0, -0.468, 0.031)$ and $(\bpsi_1,\bpsi_2,\bpsi_3) = (0, -0.465, 0.032)$. 

We estimate the nuisance functions using the ensemble method \texttt{SuperLearner} \citep{van2007super}. In estimating the conditional mean $\bm$ and the density ratio $dQ_{\bX}/dP_{\bX}$, we use a library including generalized linear models with and without interaction, random forest, and generalized additive model with degrees of freedom 3, 5 and 10. For estimating the propensity $\pi$, we substitute random forest with the overall mean. All other hyperparameters in the super learner are at their default values. For comparison, we also consider estimating these nuisance functions with working generalized linear models, with main effects only for $\pi$, main effects and interaction for $\bm$ and the density ratio, and an additional quadratic term in $X_1$ for $\bm$. Note that as the logit of $m_{12}$ is highly non-linear, the working model is misspecified.

\begin{figure}[h]
    \centering
    \includegraphics[width=0.7\textwidth,height=0.35\textwidth]{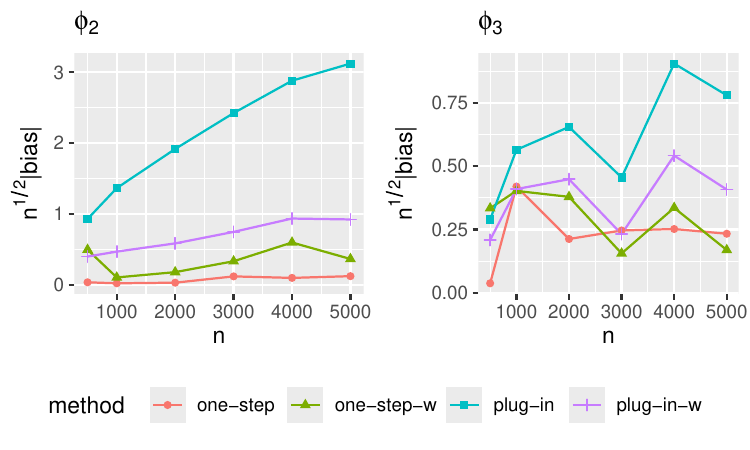}
    \includegraphics[width=0.7\textwidth,height=0.35\textwidth]{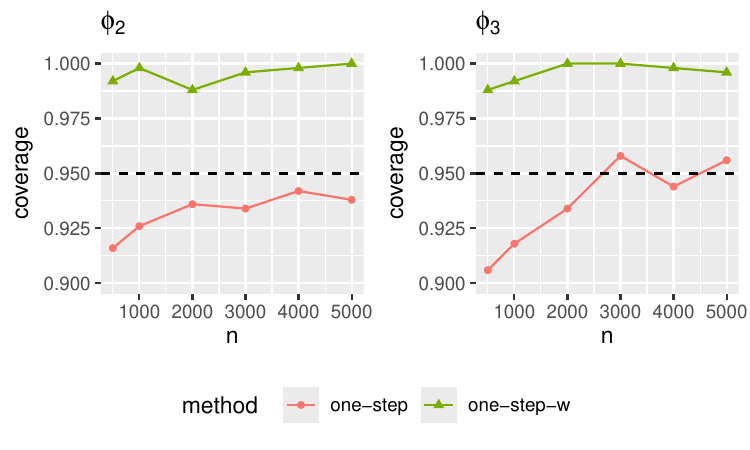}
    \caption{Estimation and inference of $\bphi$: scaled bias of plug-in and one-step estimators (upper panel) and coverage of Wald CI associated with one-step estimators (lower panel) under varying sample sizes. ``-w" indicates methods with working parametric models. Results are based on 500 simulation replications.}
    \label{fig: nonparam phi}
\end{figure}

We set the sample sizes in the labeled and unlabeled data $n=m$, and vary $n$ and $m$ in $\{500,1000,2000,3000,4000,5000\}$. We implement the estimators in Section~\ref{sec: nonparam estimation} and examine (1) the absolute bias scaled by $n^{1/2}$ of the plug-in and one-step estimators with super learner and working parametric models, and (2) the coverage of Wald confidence intervals associated with the one-step estimators. We present the results on $\bphi$ in Figure~\ref{fig: nonparam phi} and defer the results on $\bpsi$ to Appendix~\ref{app:sim}.

We observe that the absolute bias of the plug-in estimators scaled by $n^{1/2}$ is generally increasing as $n$ increases. This indicates that these plug-in estimators have a non-negligible bias for the purpose of statistical inference. The scaled bias of the one-step estimators remains stable; and for the implementation with super learner, the scaled bias appears to show a decreasing trend. As for the coverage of the Wald CI, we see that the CI associated with the one-step estimator with super learner has coverage close to the nominal level, especially with moderate sample sizes. In contrast, the CI based on working parametric models is anti-conservative, likely due to the misspecification of the conditional mean model. Results for $\bpsi$ show a similar pattern.

Next, we consider a setting (Setting II) where the data generating distribution belongs to the conditional BT model. Specifically, we let $K=5$, and set $\theta_2(\bx) = x_1x_2$, $\theta_3(\bx) = x_1^2 + x_2$, $\theta_4(\bx) = 0.5x_1 + x_2$, and $\theta_5(\bx) = \sin\{ 1.5(x_1+0.5x_2)\}$. The covariate vectors in the labeled and unlabeled data follow the same distribution as before. For $a \in \{(1,2),(2,3),(2,4),(2,5),(3,5)\}$, we let $\pi(a|\bx) = 0.2$ for all $\bx$. Under this setup, $\bphi = (0.1, 0.483, 0.525, 0.567)$, and $\bpsi = (0.091, 0.459, 0.5, 0.548)$. Here, we see a larger discrepancy between $\bphi$ and $\bpsi$ demonstrating the non-collapsibility. We implement two versions of the one-step estimators in Section~\ref{sec: cond BT estimation}: the first one is based on the influence function in Proposition~\ref{prop: IF conditional BT} where we use all the pairs except $(3,5)$ in the matrix $\Gamma$; the other one-step estimator is based on the EIF that uses information in all observed pairwise comparisons (5 in total.) The nuisance functions are estimated with the same super learner and working parametric models. Sample sizes of the labeled and unlabeled data are equal and varied in $\{2000,3000,4000,5000, 6000\}$. Results for $\phi_2$ and $\phi_5$ are presented in Figure~\ref{fig: cond phi25}, and results for $\phi_3$ and $\phi_4$ are presented in Figure~\ref{fig: cond phi34} in Appendix~\ref{app:sim}. All results for $\bpsi$ are also deferred to Appendix~\ref{app:sim}, and the patterns observed are similar to those for $\bphi$.

\begin{figure}
    \centering
    \includegraphics[width=0.7\textwidth,height=0.35\textwidth]{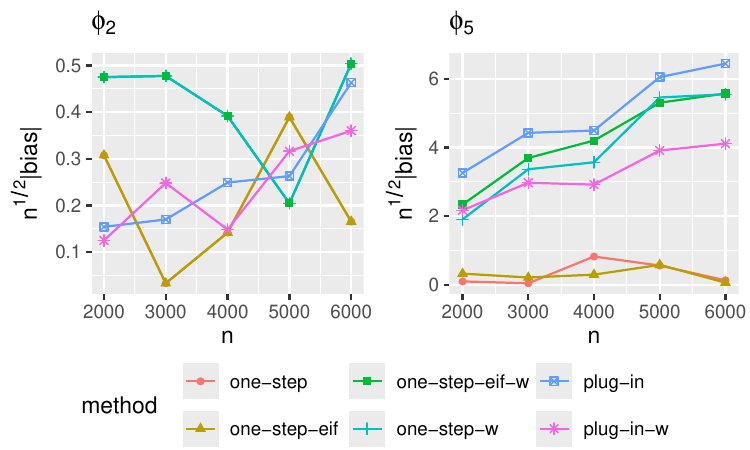}
    \includegraphics[width=0.7\textwidth,height=0.35\textwidth]{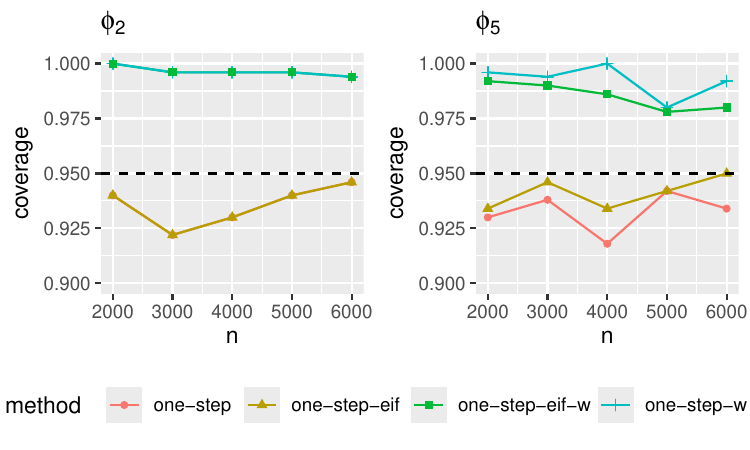}
    \includegraphics[width=0.7\textwidth,height=0.35\textwidth]{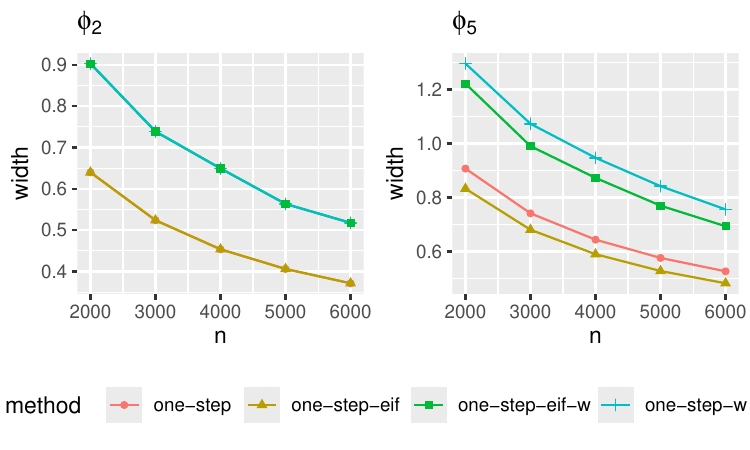}
    \caption{Estimation and inference of $\bphi_2$ and $\bphi_5$: scaled bias of plug-in and one-step estimators (upper panel), coverage (middle panel) and average width (bottom panel) of Wald CI associated with one-step estimators under varying sample sizes. ``-w" indicates methods with working parametric models. Results are based on 500 simulation replications. Data generating distribution belongs to the conditional BT model.}
    \label{fig: cond phi25}
\end{figure}

First, as the working model approximates the highly nonlinear function $\theta_5$ poorly, we see that all estimators based on the working parametric model have increasing bias when scaled properly by sample size. The plug-in estimator with nuisance functions estimated via super learner also has non-negligible bias. In contrast, the one-step estimators based on the efficient and inefficient influence functions with super learner have stable scaled bias. In terms of the coverage, we observe that the CIs associated with the one-step estimators using working models are anti-conservative and have coverage very close to 1, again likely due to model misspecification. Those based on super learner have coverage close to nominal level. The one-step estimators for $\phi_2$ and $\phi_4$ have nearly identical performances. However, for $\phi_3$ and $\phi_5$, the CI based on the efficient influence function has slightly better coverage and are narrower, demonstrating the efficiency gain. This is intuitive as the EIF-based estimator uses the additional information in the pairwise comparison between players 3 and 5.

\subsection{Evaluating the alignment of LLMs with human preference}
\label{s:data application}

We illustrate the use of the proposed approach in evaluating the alignment of LLMs with human preferences using two data sources collected from the Chatbot Arena: MT-bench and Chatbot Arena. MT-bench consists of 80 high-quality questions (Table~\ref{tab:sample_bench}) designed to access a chatbot's ability to engage in multi-turn conversations and follow instructions. Answers to these prompts were generated by six models: GPT-4, GPT-3.5, Claude-V1, Vicuna-13B, Alpaca-13B \citep{taori2023stanford}, and LLaMA-13B \citep{touvron2023llama}. Human judges, mostly graduate students, each evaluated at least 20 random multiple-turn questions, leading to around 3K pairwise votes. In addition, the Chatbot Arena crowdsourced dataset features anonymous battles between chatbots in real-world scenarios, where users engage in conversations with two chatbots and rank their responses based on personal preferences. In addition to the six aforementioned models, there were 14 others participated in the battles. Data was collected after running Chatbot Arena for a month, resulting in around 30K votes. 

Each of these two datasets has its unique strength and weakness. MT-bench consists of expert-verified labels with minimal measurement error or bias, making it well-suited for statistical estimation. In contrast, Chatbot Arena data is crowd-sourced, and the varying quality of collected votes introduces challenges in accurately assessing the underlying true performance of LLMs. As a result, one might attempt to use the MT-bench data alone for evaluting LLMs. However, it is important to point out that these carefully crafted questions in MT-bench may misrepresent the scope and types of real-world prompts, rendering MT-bench itself insufficient to reflect how LLMs are performing in practice. On the other hand, prompts collected from Chatbot Arena more closely aligns with those in real life. It is therefore of interest to evaluate the performance of LLMs using both data, where we can transport the rankings of LLMs stratified on types of prompts from the MT-bench population to our target population of interest: the real-world prompts from Chatbot Arena. 

\begin{figure}[h]
    \centering
    \includegraphics[width=0.7\linewidth]{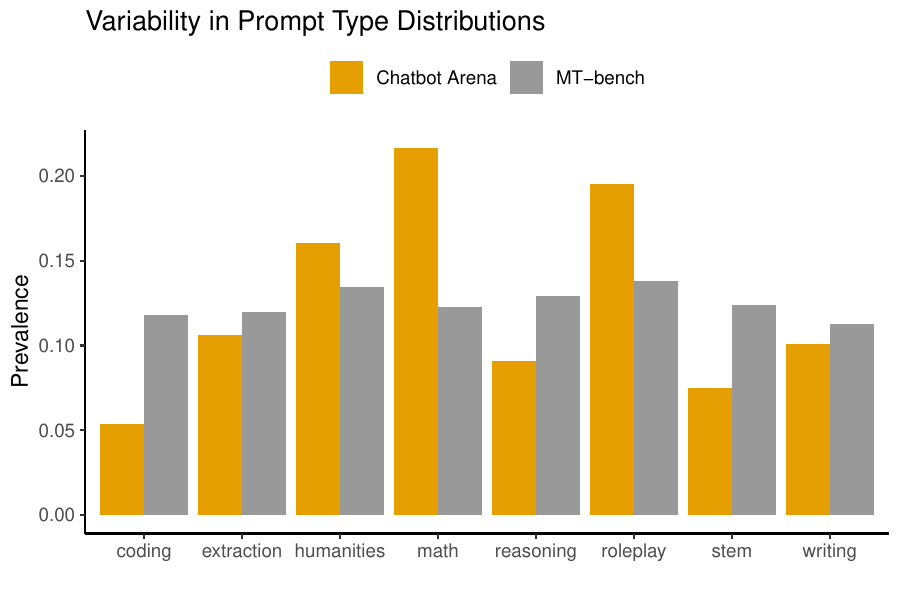} 
      \caption{Prevalence of each type of prompts in MT-bench data and Chatbot arena crowd-source data.}  \label{fig:prompt types}
\end{figure}

We aim to estimate the rankings of the six aforementioned LLMs in real-world settings using prompts sourced from the Chatbot Arena. Specifically, we discarded collected votes from Chatbot Arena and only extracted the prompts used in battles between the six aforementioned LLMs.  We used prompts from MT-bench as training data to characterize each of the prompts sourced from Chatbot Arena into one of the eight types (Table~\ref{tab:sample_arena}) using TF-IDF vectorization and a Support Vector Machine classifier. As shown in Figure~\ref{fig:prompt types}, the prompts in Chatbot Arena are more geared towards mathematics, role play and humanities, with less concerning with coding and knowledge in STEM. Between the six models, there appears to be sufficient overlap in pairwise battle counts in both MT-bench and Chatbot Arena (Figure~\ref{fig:battle_combinations}). The pairwise win fractions vary slightly between the two, possibly due to the different distributions of prompts deployed and the varying qualities of the votes.

\begin{figure}[H]
    \centering
    \includegraphics[width=0.8\linewidth]{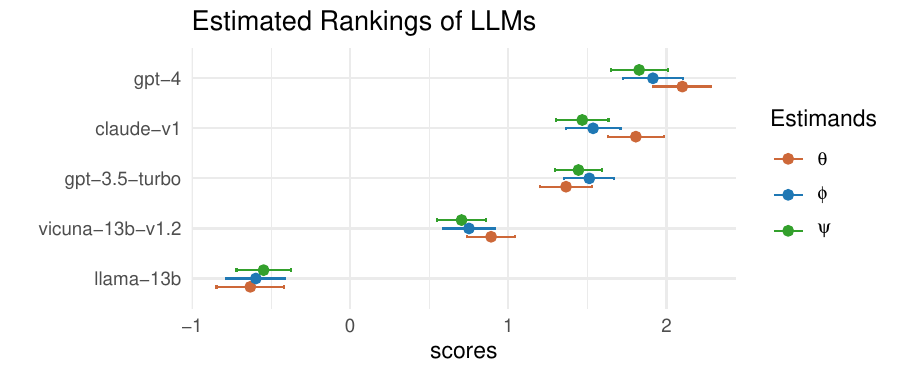}
    \caption{Estimated rankings along with 95\% CI of each model using Alpaca-13B  as the reference model. }
    \label{fig:result}
\end{figure}

We assumed that the conditional BT model is potentially misspecified, and estimated $\bphi$ and $\bpsi$ using alpaca-13b as the reference model.  In our setting, $X$ denotes the type of prompt and therefore is categorical. We treated ties as a half win, and adopted a uniform sampling scheme $\rho$. Nuisance functions were estimated via SuperLearner \citep{van2007super} with a library containing a generalized linear model with interaction terms and random forest under their default settings in the \texttt{SuperLearner} R package \citep{polley2011super}. We estimated the propensity scores using a generalized linear model with interaction terms. The density ratio of $X$ were estimated empirically. In addition to estimators of $\bphi$ and $\bpsi$ , we also report the estimated strengths $\btheta$ based on the marginal BT model fitted on Chatbot Arena using crowdsourced votes as the outcome (Figure~\ref{fig:result}). Overall, we obtained a consistent ranking of the LLMs based on metrics $\bphi$ and $\bpsi$. GPT-4 outperforms all others, while the rankings of Claude-V1 and GPT-3.5 are extremely close. By comparison, the naive estimate of $\btheta$ using crowdsourced votes gives a higher score for top players and a lower score for poor players. Such differences are possibly due to two reasons. Besides the potential misalignment of the conditional distribution of the outcome due to possibly contaminated quality of crowd-sourced votes, another main cause lies in sampling probability. In contrast with using the observed probability of sampling in marginal BT models, our approach adopts a uniform sampling scheme $\rho$ to give each pair of LLMs the same chance to be evaluated, which in turn mitigates potential bias introduced by sampling preference.

\section{Discussion and conclusion}
In this paper, we propose a general framework for statistical inference on the overall strengths of players given outcomes and contextual information of pairwise comparisons. We consider a conditional Bradley-Terry model incorporating covariates, where the strengths of the players are flexible functions of the contextual covariates. We define the estimands of interest through the Kullback-Leibler projection of any target distribution onto this model class, without relying on any stringent parametric assumptions on this target distribution. We propose efficient estimators of these estimands and develop corresponding inferential procedures in the presence of possible covariate shift. If the target distribution indeed belongs to the conditional BT model, we propose additional estimators that do not require observing all pairwise comparisons in the data. 

The BT model framework imposes the assumption that the $K(K-1)/2$ (marginal or conditional) winning probabilities in all pairwise comparisons $\bm$ are functions of a $(K-1)$-dimensional vector $\btheta$, an assumption that can be violated. Under such violation, the probabilistic limit of the BT model estimate depends on the winning probabilities in all pairwise comparisons. Hence, in general, we need to observe a fully connected comparison graph for identification and estimation. This may be challenging to achieve when the number of players $K$ is large, and the number of comparisons for a given pair may be small. To address this, we have considered correct specification of the conditional BT model to reduce the number of pairs needed for identification. But to ensure an invertible comparison matrix $\Gamma$, we cannot have disjoint components in the comparison graph and still need at least $K-1$ pairs. This highlights the importance of balancing the different pairs drawn under various contexts in study design and data collection.

The BT model is based on the logistic link function in the sense that the winning probabilities are obtained via a sigmoid transformation of the differences in players' strengths. Our proposed methods can be easily adapted to accommodate models based on other link functions. For example, replacing the logistic function with the cumulative distribution function of a standard normal distribution results in the Thurstone model \citep{thurstone2017law}. Our method can be adapted to infer overall strengths in such a model adjusting for covariates.

An alternative direction is to consider an asymptotic regime where the number of players tends to infinity under a correctly specified conditional BT model. One idea is to assume that, in addition to a group of players of interest, there are a growing number of other players drawn from a hypothetical player population. When the comparison graph is not too sparse, it may be possible to infer the overall strengths of the players of interest in a nonparametric fashion accounting for covariate information, where the length of the CI shrinks with the node degree of the player in the comparison graph. We leave this for future exploration. 

\vskip 0.2in
\bibliography{ref}

\newpage

\appendix

\section{One-step estimation of $\bpsi$}\label{app:psi}

In this section, we provide more details on one-step estimation of $\bpsi$. 

\subsection{under possibly misspecified conditional BT model}
In Proposition~\ref{prop: EIF no shift} of Section~\ref{sec: nonparam estimation}, we present the EIF of $\bpsi$ with respect to a locally nonparametric model without assuming that the conditional BT model is correctly specified. We now detail the construction of the corresponding one-step estimator. First, given the estimates of the conditional mean functions $\widehat m_{kl}(\bx)$, we can estimate all the marginal winning probabilities $m_{kl} = q_{kl}$ by $\widehat m_{kl} = n^{-1}\sum_{i=1}^n \widehat m_{kl}(\bX_i)$. We can then solve the equation in \eqref{eq: marginal EE} to obtain a plug-in estimator of $\bpsi$, denoted as $\widetilde\bpsi$ such that
\begin{equation}\label{eq: plug in psi}
    \sum_{l \neq k} \rho_{kl}\left\{\sigma\left(\widetilde\psi_k - \widetilde\psi_l\right) - \widehat m_{kl}\right\} = 0.
\end{equation}
The one-step estimator $\widehat\bpsi$ is constructed as follows:
\begin{equation*}
    \widehat\bpsi = \widetilde\bpsi - n^{-1}\sum_{i=1}^n \widehat\Lambda(\widetilde\bpsi,\widehat\bm)\left\{\widehat\tau(\bX_i,A_i,Y_i) + \widehat\bm(\bX_i) - \widehat\bm\right\},
\end{equation*}
where $\widehat\tau$ is defined in the same way as $\tau$ but with all nuisance functions replaced by their estimates. In the presence of covariate shift, we estimate $\bpsi$ via data fusion. The corresponding EIF is given in Proposition~\ref{prop: EIF with shift}. Again, given the nuisance function estimates we construct plug-in estimators of $\bm$ by marginalizing over the empirical distribution of $\bX$ in the target data, $\widehat m_{kl} = m^{-1}\sum_{i=1}^{n+m}S_i \widehat m_{kl}(\bX_i)$. We then define the plug-in estimator $\widetilde\bpsi_f$ in the same way as in \eqref{eq: plug in psi}, and a one-step estimator as follows
\begin{equation*}
    \widehat\bpsi_f = \widetilde\bpsi_f - N^{-1}\sum_{i=1}^N \widehat\Lambda(\widetilde\bpsi_f,\widehat\bm) \left\{\frac{S_i}{\bar S}\widehat w(X_i)\widehat\tau(\bX_i,A_i,Y_i) + \frac{(1-S_i)}{1-\bar S}\left(\widehat\bm(\bX_i) - \widehat\bm\right)\right\},
\end{equation*}
where $\bar S$ is the sample mean of $S_i$ and $N = n+m$.

\subsection{under correctly specified conditional BT model}
When the conditional BT model is indeed correctly specified, we can identify $\btheta(\bx)$ with fewer pairwise comparisons. Since the conditional winning probabilities in all pairwise comparisons $\bm(\bx)$ are fully parametrized by $\btheta(\bx)$, $\bm(\bx)$ itself can be identified and so are $\bm$ and $\bpsi$. Specifically, $m_{kl}(\bx) = \sigma(\theta_k(\bx) - \theta_l(\bx))$, $\bm(\bx)$ is the concatenation of $m_{kl}(x)$, and $\bm = \EE_Q[\bm(\bX)]$.

\begin{proposition}[IF of $\bPsi$ under correctly specified conditional BT model]\label{prop: IF psi cond BT}
Suppose that $Q \in \mathcal{P}$. Under Assumptions~\ref{cond: no shift cond} and \ref{cond: gamma full rank}, the following functions are an influence function of and the EIF of $\bPsi$:
\begin{align*}
    D_{cond,\bpsi}: (\bx,a,y) &\mapsto -\Lambda(\bpsi,\bm)\left\{\frac{\partial\bm(\bx)}{\partial\btheta(\bx)} \Gamma^{-1}\tilde\tau(\bx,a,y) + \bm(\bx) - \bm\right\}; \\
    D_{cond,\bpsi}^*: (\bx,a,y) &\mapsto -\Lambda(\bpsi,\bm)\left\{\frac{\partial\bm(\bx)}{\partial\btheta(\bx)} \left(\Gamma_*^\top W_*(\bx) \Gamma_*\right)^{-1}\Gamma_*^\top v(\bx,a,y) + \bm(\bx) - \bm\right\}.
\end{align*}
Under Assumptions~\ref{cond: no shift cond}(b)-(c), \ref{cond: with shift cond} and \ref{cond: gamma full rank}, the following functions are an influence function of and the EIF of $\bPsi_f$ in the data fusion setting:
\begin{multline*}
    D_{cond,f,\bpsi}: (s,\bx,a,y) \mapsto -\Lambda(\bpsi,\bm)\Big\{\frac{s}{Pr(S=1)}\frac{dQ_{\bX}}{dP_{\bX}}(\bx)\frac{\partial\bm(\bx)}{\partial\btheta(\bx)} \Gamma^{-1}\tilde\tau(\bx,a,y) \\
    + \frac{1-s}{Pr(S=0)}\left(\bm(\bx) - \bm\right)\Big\}.
\end{multline*}
\begin{multline*}
    D_{cond,f,\bpsi}^*: (s,\bx,a,y) \mapsto -\Lambda(\bpsi,\bm)\Big\{\frac{s}{Pr(S=1)}\frac{dQ_{\bX}}{dP_{\bX}}(\bx)\frac{\partial\bm(\bx)}{\partial\btheta(\bx)} \left(\Gamma_*^\top W_*(\bx) \Gamma_*\right)^{-1}\Gamma_*^\top v(\bx,a,y) \\
    + \frac{1-s}{Pr(S=0)}\left(\bm(\bx) - \bm\right)\Big\}.
\end{multline*}
\end{proposition}
The function $\tilde\tau$ is defined in Proposition~\ref{prop: IF conditional BT} in Section~\ref{sec: cond BT estimation}, and the explicit expression of the partial derivative $\partial\bm(\bx)/\partial\btheta(\bx)$ is given in the proof of this result. 

Given a comparison matrix $\Gamma$ for identification and initial estimates of the winning probabilities of pairs in $\Gamma$, $\widehat\bm_\Gamma(\bx)$, we estimate $\btheta(\bx)$ via $\widehat\btheta(\bx) = \Gamma^{-1}\sigma^{-1}(\widehat\bm_\Gamma(\bx))$. Consequently, we can construct a plug-in estimator of $\bm$ --- without covariate shift $\widehat\bm = n^{-1}\sum_{i=1}^{n}\widehat\bm(\bX_i)$; and in the data fusion setting $\widehat\bm = m^{-1}\sum_{i=1}^{n+m}S_i \widehat\bm(\bX_i)$ --- both with $\widehat m_{kl}(\bx) = \sigma(\widehat\theta_k(\bx) - \widehat\theta_l(\bx))$. Plug-in estimators $\widetilde\bpsi$ and $\widetilde\bpsi_f$ can be constructed via \eqref{eq: plug in psi}, and corresponding one-step estimators are
\begin{align*}
    \widehat\bpsi &= \widetilde\bpsi - n^{-1}\sum_{i=1}^n \widehat\Lambda(\widetilde\bpsi,\widehat\bm)\left\{\frac{\partial\widehat\bm(\bX_i)}{\partial\widehat\btheta(\bX_i)} \Gamma^{-1}\tau^\ddagger(\bX_i,A_i,Y_i) + \widehat\bm(\bX_i) - \widehat\bm\right\}.\\
    \widehat\bpsi_f &= \widetilde\bpsi_f - N^{-1}\sum_{i=1}^N  \widehat\Lambda(\widetilde\bpsi_f,\widehat\bm)\left\{\frac{S_i}{\bar S}\widehat w(X_i)\frac{\partial\widehat\bm(\bX_i)}{\partial\widehat\btheta(\bX_i)} \Gamma^{-1}\tau^\ddagger(\bX_i,A_i,Y_i) + \frac{1-S_i}{1-\bar S}\left(\widehat\bm(\bX_i) - \widehat\bm\right)\right\},
\end{align*}
where $\tau^\ddagger$ is defined in the same way as $\tilde\tau$ but with all unknown nuisance functions replaced by their corresponding estimates. The estimators based on the EIF can be defined similarly, where we replace $\Gamma^{-1}\tau^\ddagger(\bX_i,A_i,Y_i)$ with $(\Gamma_*^\top \widehat W_*(\bX_i) \Gamma_*)^{-1}\Gamma_*^\top \widehat{v}(\bX_i,A_i,Y_i)$.

\section{Proofs of results in Section~\ref{sec: cond BT estimation} and additional theoretical results under conditional BT model}\label{app:graph}

\subsection{Full column rank of $\Gamma$}\label{subapp: connected graph}

For the ease of presentation, in defining the comparison matrix $\Gamma \in \mathbb{R}^{J\times (K-1)}$, we left out the column corresponding to player 1. We can define a similar matrix $\widetilde{\Gamma} \in \mathbb{R}^{J\times K}$ by adding that column to $\Gamma$. Specifically, each row of $\widetilde{\Gamma}$ contains exactly one 1 and one -1, and define $k_j < l_j$ such that $\widetilde{\Gamma}_{j,k_j} =1$ and $\widetilde{\Gamma}_{j,l_j} = -1$ for all $1\leq j \leq J$. The $j$-th row therefore represents a comparison between player $k_j$ and $l_j$. We can use the matrix $\widetilde{\Gamma}$ as an incidence matrix and form an undirected graph $G(\widetilde\Gamma)$ with node set $\{1,\ldots,K\}$, which we refer to as the comparison graph. One example is given below for 5 players.
$$
\Gamma = 
\begin{bmatrix}
    -1& 0 & 0 & 0 \\
    0 & 1 & -1 & 0 \\
    0 & 1 & 0 & -1 \\
    0 & 0 & 1 & -1 \\
    1 & 0 & 0 & -1 
\end{bmatrix}
\quad
\widetilde\Gamma = 
\begin{bmatrix}
    1 & -1& 0 & 0 & 0 \\
    0 & 0 & 1 & -1 & 0 \\
    0 & 0 & 1 & 0 & -1 \\
    0 & 0 & 0 & 1 & -1 \\
    0 & 1 & 0 & 0 & -1 
\end{bmatrix}
$$

\begin{figure}[h]
\centering
\begin{tikzpicture}[node distance={15mm}, main/.style = {draw, circle}] 
\node[main] (1) {$1$}; 
\node[main] (2) [above left of=1] {$2$};
\node[main] (5) [above right of=1] {$5$}; 
\node[main] (3) [below right of=5] {$3$};
\node[main] (4) [above right of=3] {$4$};
\draw (1) -- (2);
\draw (2) -- (5);
\draw (3) -- (4);
\draw (3) -- (5);
\draw (4) -- (5);
\end{tikzpicture} 
\caption{A comparison graph corresponding to $\Gamma$ and $\widetilde\Gamma$}
\end{figure}
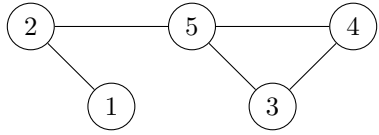

\begin{proposition}\label{prop: connected graph}
    $G(\widetilde\Gamma)$ is connected if and only if Assumption~\ref{cond: gamma full rank} holds, that is, when $\Gamma$ has full column rank.
\end{proposition}

\begin{proof}
Let $\bOne$ denote the vector with all 1s of appropriate dimension. Since each row of $\widetilde\Gamma$ sums up to 0, the vector $\bOne$ is in the null space of $\widetilde\Gamma$. 

If Assumption~\ref{cond: gamma full rank} holds, we know that $\widetilde\Gamma$ has at least $K-1$ linearly independent columns as the last $K-1$ columns are the same columns in $\Gamma$ and are linearly independent. Therefore, rank of $\widetilde\Gamma$ is $K-1$ and by Proposition 4.3 in \cite{biggs1993algebraic}, $G(\widetilde\Gamma)$ has one connected component. 

To show the reverse, if $G(\widetilde\Gamma)$ is connected, $\widetilde\Gamma$ has rank $K-1$ by Proposition 4.3 in \cite{biggs1993algebraic}. As the null space is spanned by the vector $\bOne$, removing any column would lead to a matrix with full column rank. 
\end{proof}

Proposition~\ref{prop: connected graph} thus offers an intuitive interpretation of the identifiability condition under correctly-specified conditional BT model: to identify $\theta_k$ for all $k$, we need to have pairwise comparisons such that any two players can be connected by a path.

\subsection{Tangent space under the conditional BT model and the EIF}

Recall that $\calP$ is the conditional BT model. Define a submodel $\calP_0 \subseteq \calP$ where all distributions in $\calP_0$ have the same conditional distribution of $A|\bX$ as $P_{A|\bX}$. Further suppose that $P \in \calP_0$ such that the conditional BT model is correctly specified and $P(Y=1 | A=(k,l),\bX=\bx) = \sigma(\theta_k(\bx) - \theta_l(\bx))$. The following proposition derives the tangent space at $P$ relative to the model $\calP_0$.

\begin{proposition}[Tangent space under conditional BT model]\label{prop: tangent space cond BT}
Suppose that $P(Y=1 | A=(k,l),\bX=\bx) = \sigma(\theta_k(\bx) - \theta_l(\bx))$ for $1 \leq k < l \leq K$, then the tangent space at $P$ relative to the model $\calP_0$, $\mathcal{T}$, takes the following form.
\begin{multline*}
    \mathcal{T} = \overline{\textnormal{span}}\Bigg\{f: f(\bx,a,y) =  \sum_{1 \leq k<l \leq K}I\{a=(k,l)\} I\{\pi((k,l)|\bx) > 0\}\left\{y-\sigma(\theta_k(\bx) - \theta_l(\bx))\right\}\left\{h_k(\bx) - h_l(\bx)\right\}\\
    + h(\bx) \textnormal{ for some} \ h \in L^2_0(P_{\bX})\ \textnormal{and} \ \{h_k: \ 1\leq k\leq K, h_1 = \boldsymbol{0}\}, \ f \in L^2_0(P) \Bigg\}.
\end{multline*}
\end{proposition}

\begin{proof}[Proof of Proposition~\ref{prop: tangent space cond BT}]
Under a correctly specified conditional BT model, the condtional distributions $P_{Y|(A=a,X)}$ for different values of $a$ are no longer variational independent.

Consider the following perturbed likelihood
\begin{align*}
    \log p_\epsilon(\bx,a,y) &= \log \{p_{\bX}(\bx)(1+\epsilon h(\bx))\} + \sum_{1 \leq k < l \leq K}I\{a=(k,l)\}I\{\pi((k,l)|\bx) > 0\}\log \pi((k,l)|\bx) \\
    &\quad + \sum_{1 \leq k < l \leq K}I\{a=(k,l)\}I\{\pi((k,l)|\bx) > 0\} y \log \sigma\left(\theta_k(\bx) + \epsilon h_k(\bx)-\theta_l(\bx) - \epsilon h_l(\bx)\right) \\
    &\quad + \sum_{1 \leq k < l \leq K} I\{a = (k,l)\}I\{\pi((k,l)|\bx) > 0\} (1-y)\log \left(1-\sigma\left(\theta_k(\bx) + \epsilon h_k(\bx)-\theta_l(\bx) - \epsilon h_l(\bx)\right)\right),
\end{align*}
where $h_k$ is the perturbation to $\theta_k(\cdot)$ for $k = 2,\ldots, K$ and $h_1$ is set to the zero function. Here, we use the convention that $0\log 0 = 0$. Taking derivative with respect to $\epsilon$, we get
\begin{align*}
    \frac{\partial \log p_\epsilon}{\partial \epsilon}\Bigr\rvert_{\epsilon=0} &= h(\bx) + \sum_{1 \leq k<l \leq K}I\{a=(k,l)\}I\{\pi((k,l)|\bx) > 0\}\left\{y-\sigma(\theta_k(\bx) - \theta_l(\bx))\right\}\left\{h_k(\bx) - h_l(\bx)\right\}.
\end{align*}
The closure of the linear span of the functions taking the form above is the tangent space. 
\end{proof}

Now let us define a comparison matrix $\Gamma_*$ that lists all possible pairwise comparisons between two players. That is, $\Gamma_* \in \mathbb{R}^{\frac{(K-1)K}{2} \times (K-1)}$ such that for any $a = (k,l) \in \calA$, there exists $1 \leq j \leq K(K-1)/2$ such that $k_{*j} = k$ and $l_{*j} = l$.  Here, $k_{*j}$ and $l_{*j}$ again indicate the location of $1$ and $-1$ in the $j$-th row of $\Gamma_*$ (specifically $(\Gamma_*)_{j,k_{*j}-1 } = 1$ and $(\Gamma_*)_{j,l_{*j}-1} = -1$), thus corresponding to the two players in the $j$-th pairwise comparison. Define a weight matrix $W_*(\bx)$ that is a $K(K-1)/2$ by $K(K-1)/2$ diagonal matrix, with the $j$-th diagonal element equal to $m_{k_{*j}l_{*j}}(\bx)(1-m_{k_{*j}l_{*j}}(\bx))\pi((k_{*j},l_{*j})|\bx)$ and zeros off-diagonal. The reduced Laplacian matrix of the weighted graph is $L(\bx) = \Gamma_*^\top W_*(\bx) \Gamma_*$. (Note that if the $j$-th row in $\Gamma_*$ corresponds to a pairwise comparison that is not observed, $\pi((k_{*j},l_{*j})|\bx) = 0$. This has the same effect in computing the graph Laplacian as if we remove that row from $\Gamma_*$.
\\
\\
\begin{proof}[Proof of Proposition~\ref{prop: EIF cond}]
We first derive the EIF of $\phi_i$ relative to the model $\calP_0$ by projecting a gradient onto the tangent space derived in Proposition~\ref{prop: tangent space cond BT}, for given $i \in \{2,\ldots,K\}$. To start, we note that the following function is a gradient of $\bphi$ as shown in Proposition~\ref{prop: IF conditional BT}:
\begin{equation*}
    D_{cond}: (\bx,a,y) \mapsto \Gamma^{-1}\tilde\tau(\bx,a,y) + \btheta(\bx) - \bphi,
\end{equation*}
where $\tilde\tau(\bx,a,y) = (\tilde\tau_1(\bx,a,y),\ldots, \tilde\tau_J(\bx,a,y)) \in \mathbb{R}^J$ and 
\begin{equation*}
   \tilde\tau_j(\bx,a,y) = \tau_{k_jl_j}(\bx,a,y)m_{k_jl_j}(\bx)^{-1}(1-m_{k_jl_j}(\bx))^{-1}; 
\end{equation*}
\begin{equation*}
    \tau_{kl}(\bx,a,y) =  (-1)^{I\{k < l\}+1} \frac{I\left\{a = (k\wedge l, k\vee l)\right\}}{\pi((k\wedge l, k\vee l)|\bx)} \left\{y - m_{(k\wedge l)(k\vee l)}(\bx)\right\}. 
\end{equation*}
We focus on the $(i-1)$-th component of $D_{cond}$, which is a gradient for $\phi_i$. The pseudo-inverse of $\Gamma$ is $(\Gamma^\top\Gamma)^{-1}\Gamma^\top \in \mathbb{R}^{(K-1)\times J}$, and we write the $(i-1)$-th row as $(\gamma_1^i,\ldots,\gamma_J^i)$. Using this notation, we can now unpack the expression of $D_{cond}$, and specifically,
\begin{equation*}
    D_{cond,i}: (\bx,a,y) \mapsto \sum_{j=1}^J \frac{ \gamma_j^i I\{a = (k_j,l_j)\}}{m_{k_jl_j}(\bx)(1-m_{k_jl_j}(\bx))\pi((k_j,l_j)|\bx)}\left\{y - m_{k_j l_j}(\bx)\right\}
    + \theta_i(\bx) - \phi_i.
\end{equation*}

To project onto the tangent space, we first note that $\theta_i(\bx) - \phi_i$ is an integrable mean-zero function of $\bx$ and thus is already in the tangent space. Therefore, we only need to project the first piece, which has mean 0 conditional on $a,\bx$, onto the following space:
\begin{equation*}
    \left\{\sum_{1 \leq k<l \leq K}I\{a=(k,l)\}I\{\pi((k,l)|\bx) > 0\}\left\{y-\sigma(\theta_k(\bx) - \theta_l(\bx))\right\}\left\{h_k(\bx) - h_l(\bx)\right\}, h_1 = 0\right\}.
\end{equation*}
We define a new set of weights $\{\gamma_{kl}^i: 1\leq k < l \leq K\}$ such that $\gamma_{kl}^i = \gamma_j^i$ if $k=k_j, l=l_j$ for some $j$ and $\gamma_{kl}^i = 0$ otherwise. The function we wish to project is now equivalent to
\begin{equation*}
    \sum_{1 \leq k < l \leq K} \frac{\gamma_{kl}^i I\left\{a = (k,l)\right\}}{\pi((k,l)|\bx)m_{kl}(\bx)(1-m_{kl}(\bx))} \left\{y - m_{kl}(\bx)\right\}.
\end{equation*}
To find the projection, we need to solve for a set of functions $h_k^*(\bx), 2 \leq k \leq K$ ($h_1^* = 0$) such that the following residual is orthogonal to any function in the tangent space.
\begin{equation*}
    \sum_{1\leq k < l \leq K} \left(\frac{\gamma_{kl}^i}{\pi((k,l)|\bx)m_{kl}(\bx)(1-m_{kl}(\bx))} - I\{\pi((k,l)|\bx) > 0\}\left\{h_k^*(\bx) - h_l^*(\bx)\right\} \right) I\{a=(k,l)\}\left\{y-\sigma(\theta_k(\bx) - \theta_l(\bx))\right\},
\end{equation*}
which means that the following inner product is 0 for any $h_k, 1 \leq k \leq K$.
\begin{multline*}
    \EE\Bigg[\sum_{1\leq k < l \leq K} \left(\frac{\gamma_{kl}^i}{\pi((k,l)|\bx)m_{kl}(\bx)(1-m_{kl}(\bx))} - I\{\pi((k,l)|\bx) > 0\}\left\{h_k^*(\bx) - h_l^*(\bx)\right\} \right) \times \\
    I\{a=(k,l)\} I\{\pi((k,l)|\bx) > 0\} \left\{y-\sigma(\theta_k(\bx) - \theta_l(\bx))\right\}^2\left\{h_k(\bx) - h_l(\bx)\right\}\Bigg] = 0
\end{multline*}
Applying the tower property, we get
\begin{equation*}
    \EE\left[\sum_{1\leq k < l \leq K} \left(\gamma_{kl}^i - \pi((k,l)|\bx)m_{kl}(\bx)(1-m_{kl}(\bx))\left\{h_k^*(\bx) - h_l^*(\bx)\right\} \right) 
    \left\{h_k(\bx) - h_l(\bx)\right\}\right] = 0
\end{equation*}
For notational convenience, define $g_{kl}(\bx) = \pi((k,l)|\bx)m_{kl}(\bx)(1-m_{kl}(\bx))$, for $1\leq k < l \leq K$; and $g_{lk}(\bx) = g_{kl}(\bx)$. 

Now fix an index $2 \leq j \leq K$, and set all other $h_k$ to the zero function, we get
\begin{align*}
    0 &= \EE\left[\sum_{k < j } \left(\gamma_{kj}^i - g_{kj}(\bx)\left\{h_k^*(\bx) - h_j^*(\bx)\right\} \right)(-h_j(\bx)) + \sum_{j<k} \left(\gamma_{jk}^i - g_{jk}(\bx)\left\{h_j^*(\bx) - h_k^*(\bx)\right\} \right) h_j(\bx)\right] \\
    &= \EE\left[\left( - \sum_{k < j } \left(\gamma_{kj}^i - g_{kj}(\bx)\left\{h_k^*(\bx) - h_j^*(\bx)\right\} \right) + \sum_{j<k} \left(\gamma_{jk}^i - g_{jk}(\bx)\left\{h_j^*(\bx) - h_k^*(\bx)\right\} \right) \right) h_j(\bx)\right].
\end{align*}
As $h_j(\bx)$ is an integrable but otherwise unrestricted function, we must have
\begin{equation*}
    \sum_{k < j } \left( -\gamma_{kj}^i - g_{kj}(\bx)\left\{h_j^*(\bx) - h_k^*(\bx)\right\} \right) + \sum_{j<k} \left(\gamma_{jk}^i - g_{jk}(\bx)\left\{h_j^*(\bx) - h_k^*(\bx)\right\} \right) = 0,
\end{equation*}
or equivalently
\begin{equation}\label{eq: equation for h star}
    -\sum_{k<j} g_{kj}(\bx)h_k^*(\bx) + \left(\sum_{k<j}g_{kj}(\bx) + \sum_{j<k} g_{jk}(\bx)\right)h_j^*(\bx) - \sum_{j<k} g_{jk}(\bx)h_k^*(\bx)= - \sum_{k<j} \gamma_{kj}^i + \sum_{j<k} \gamma_{jk}^i.
\end{equation}
We will show that the right-hand side of the above equation simplies to $I\{i = j\}$. To see this, let consider $- \sum_{b<a} \gamma_{ba}^i + \sum_{a<b} \gamma_{ab}^i$ for a fixed $a \geq 2$. Note that by construction, $\gamma_{ab}$ is non-zero only if $a = k_j$ for some $j$ and $b = l_j$; and similarly, $\gamma_{ba}$ is non-zero only if $a = l_j$ for some $j$ and $b = k_j$. Therefore
\begin{equation*}
    - \sum_{b<a} \gamma_{ba}^i + \sum_{a<b} \gamma_{ab}^i = \sum_{j=1}^J I\{a = k_j\}\gamma_{k_jl_j}^i - \sum_{j=1}^J I\{a = l_j\}\gamma_{k_jl_j}^i = \sum_{j=1}^J \gamma_{k_j l_j}^i \left(I\{a = k_j\} - I\{a = l_j\}\right).
\end{equation*}
The $j$-th entry in the $(a-1)$-th column of $\Gamma$ is -1 if $l_j = a$ and 1 if $k_j = a$, which is equivalent to $I\{a = k_j\} - I\{a = l_j\}$ since $k_j \neq l_j$. Therefore, the expression in the display above is the inner product between the $(i-1)$-th row of $(\Gamma^\top\Gamma)^{-1}\Gamma^\top$ and the $(a-1)$-th column of $\Gamma$. Since $(\Gamma^\top\Gamma)^{-1}\Gamma^\top\Gamma$ is the identity matrix, we get $- \sum_{b<a} \gamma_{ba}^i + \sum_{a<b} \gamma_{ab}^i = I\{a = i\}$. Going back to \eqref{eq: equation for h star}, we get
\begin{equation*}
    -\sum_{k<j} g_{kj}(\bx)h_k^*(\bx) + \left(\sum_{k<j}g_{kj}(\bx) + \sum_{j<k} g_{jk}(\bx)\right)h_j^*(\bx) - \sum_{j<k} g_{jk}(\bx)h_k^*(\bx)= I\{j = i\}.
\end{equation*}

The above argument applies to any fixed $j \in \{2,\ldots,K\}$, and hence we get $K-1$ such equations, one for each $j$. This defines a system of linear equations in $\bh^*(\bx) = (h_2^*(\bx),\ldots,h_K^*(\bx))$ for given $\bx$.
$$
\begin{bmatrix}
    \sum_{k\neq 2} g_{2k}(\bx) & -g_{23}(\bx) & \hdots & -g_{2K}(\bx) \\
    -g_{23}(\bx) & \sum_{k\neq 3} g_{3k}(\bx) & \hdots & -g_{3K}(\bx) \\
    \vdots & \vdots & \ddots & \vdots \\
    -g_{2K}(\bx) & -g_{3K}(\bx) & \hdots & \sum_{k \neq K} g_{Kk}(\bx)
\end{bmatrix}
\begin{bmatrix}
    h_2^*(\bx) \\
    h_3^*(\bx) \\
    \vdots \\
    h_K^*(\bx)
\end{bmatrix}
=
\begin{bmatrix}
    I\{i=2\} \\
    I\{i=3\} \\
    \vdots \\
    I\{i=K\}
\end{bmatrix}
$$
This system of equations alone will uniquely determine $\bh^*(\bx)$ if the first matrix, denoted by $L(\bx)$, is invertible. We will now show that this is indeed the case.

Using yet another connection to graph theory, we see that $L(\bx)$ is in fact the principal submatrix of a Laplacian matrix of a graph, obtained by removing the first row and first column of the Laplacian matrix. The graph $G(L(\bx))$ is a weighted graph with edge weight $g_{kl} = g_{lk} = \pi((k,l)|\bx)m_{kl}(\bx)(1-m_{kl}(\bx))$ between nodes $k < l$. When the conditional mean $m_{kl}(\bx)$ is bounded away from 0 and 1 for all $(k,l)$, we see that an edge exists between two nodes if and only if $\pi((k,l)|\bx) > 0$. Under Assumptions~\ref{cond: no shift cond}(c) and \ref{cond: gamma full rank}, $\pi(a|\bx) > \delta$ for all $a \in \calA_J$ and this set of edges alone make $G(L(\bx))$ a connected graph (see Proposition~\ref{prop: connected graph}). By the weighted matrix tree theorem \citep{klee2020linear}, $L(\bx)$ is invertible. Alternatively, the invertibility can be established by the fact that a weakly chained diagonally dominated matrix is non-singular \citep{shivakumar1974sufficient}. It is straightforward to see that $L(\bx)$ is weakly diagonally dominated and that it has at least one strictly diagonally dominated (SDD) row (the row where $\pi((1,k)|\bx) > \delta$, which must exist because the graph is connected and 1 is connected to some other node.) Next, we show that for a non-SDD row, there is a path to a SDD row. This is straightforward if $G(L(\bx))$ remains connected after removing node 1. If removing node 1 disconnects the graph and results in more than 1 connected components, there must be a node in each connected component that is also connected to node 1 in the original graph $G(L(\bx))$, which means that there is a SDD row in each connected component. 

Solving the linear system, we get $\bh^*(\bx) = L^{-1}_{.,i-1}(\bx)$ where the subscript $_{.,i-1}$ indicates the $(i-1)$-th column of $L^{-1}(\bx)$, and $h_1^* = \boldsymbol{0}$. With this, the canonical gradient of $\phi_i: \calP_0 \rightarrow \mathbb{R}$ is
\begin{align*}
    D_{cond,i}^*: (\bx,a,y) &\mapsto \sum_{1 \leq k<l \leq K}I\{a=(k,l)\}I\{\pi((k,l)|\bx) > 0\}\left\{y-\sigma(\theta_k(\bx) - \theta_l(\bx))\right\}\left\{h_k^*(\bx) - h_l^*(\bx)\right\} \\
    &\quad + \theta_i(\bx) - \phi_i.
\end{align*}
Importantly, we observe that this EIF does not depend on $\Gamma$, the user-specified identification strategy that determines the initial gradient (if multiple identification strategies exist.) Using the notation of the full comparison matrix $\Gamma_*$, we get
\begin{align*}
    D_{cond,i}^*(\bx,a,y) &= \sum_{j=1}^{K(K-1)/2} I\{a=(k_{*j},l_{*j})\}I\{\pi((k_{*j},l_{*j})|\bx) > 0\}\left\{y-m_{k_{*j}l_{*j}}(\bx)\right\}\left\{h_{k_{*j}}^*(\bx) - h_{l_{*j}}^*(\bx)\right\} \\
    &\quad + \theta_i(\bx) - \phi_i \\
    &= v(\bx,a,y)^\top \Gamma_* \bh^*(\bx) + \theta_i(\bx) - \phi_i\\
    &= v(\bx,a,y)^\top \Gamma_* L^{-1}(\bx)e_{i-1} + \theta_i(\bx) - \phi_i \\
    &= v(\bx,a,y)^\top \Gamma_* \left(\Gamma_*^\top W_*(\bx) \Gamma_*\right)^{-1} e_{i-1} + \theta_i(\bx) - \phi_i,
\end{align*}
where we note that $L(\bx) = \Gamma_*^\top W_*(\bx) \Gamma_*$.

We now argue that $D_{cond,i}^*$ is also the canonical gradient of $\phi_i: \calP \rightarrow \mathbb{R}$. Compared to $\calP_0$, $\calP$, the conditional BT model, no longer restricts the conditional distribution of $A|\bX$. Additional score functions now arise from the perturbation of $P_{A|\bX}$. However, the canonical gradient is orthogonal to such scores that arise from perturbations that change $P_{A|\bX}$ only but leave $\btheta$ unchanged, as the parameter is independent of $P_{A|\bX}$.
Therefore, $D_{cond,i}^*$ remains the canonical gradient.

Applying the results in \cite{li2023efficient} yields the canonical gradient in the data fusion setting.
\end{proof}

Unfortunately, we are unable to give an explicit, easy-to-use expression of the matrix $L^{-1}(\bx)$ in the most general case. However, we do want to discuss some important special cases.

\noindent\textbf{Case I: minimal number of observed pairs.} If we only observe $K-1$ pairs, the minimal number needed to ensure that $\Gamma$ has full column rank, Corollary~\ref{corollary: minimal pairs} implies that the influence functions in Proposition~\ref{prop: IF conditional BT} are in fact efficient. Recall that the set $\calA_J$ denotes the set of observed pairs. For $a \in \calA_J$, positivity assumption holds.
\begin{assumption}[Minimal pairs observed]\label{cond: no additional pairs}
    $J = K-1$ and $P(A = a|\bX = \bx) = 0$ for all $\bx \in \mathcal{X}$ for all $a \notin \calA_J$.
\end{assumption}

\begin{proof}[Proof of Corollary~\ref{corollary: minimal pairs}]
Going back to Section~\ref{subapp: connected graph}, we have an incidence matrix $\widetilde\Gamma \in \mathbb{R}^{(K-1) \times K}$. Define a diagonal weight matrix $W(\bx) \in \mathbb{R}^{(K-1)\times (K-1)}$, such that $W(\bx)_{j,j} = \pi((k_j,l_j)|\bx)m_{k_jl_j}(\bx)(1-m_{k_jl_j}(\bx))$ for $1 \leq j \leq J=K-1$. The matrix $L(\bx) = (\widetilde\Gamma^\top W(\bx) \widetilde\Gamma)_{-1,-1}$ where the subscript $_{-1,-1}$ indicates removing the first row and first column, as the Laplacian matrix is $\widetilde\Gamma^\top W(\bx) \widetilde\Gamma$. $\Gamma$ is the matrix obtained by removing the first column of $\widetilde\Gamma$, and thus $L(\bx) = (\Gamma^\top W(\bx) \Gamma)$ and $L^{-1}(\bx) = \Gamma^{-1} W^{-1}(\bx)(\Gamma^\top)^{-1}$ as $\Gamma$ itself is invertible.

In this special case, the expression for the canonical gradient simplifies to
\begin{equation*}
    D_{cond,i}^*: (\bx,a,y) \mapsto \sum_{1 \leq j \leq K-1}I\{a=(k_j,l_j)\}\left\{y-m_{k_jl_j}(\bx)\right\}\left\{h_{k_j}^*(\bx) - h_{l_j}^*(\bx)\right\} + \theta_i(\bx) - \phi_i.
\end{equation*}
Define a $(K-1)$-dimensional vector $v(\bx,a,y)$ such that the $j$-th component is $I\{a=(k_j,l_j)\}\{y-m_{k_jl_j}(\bx)\}$ for $1\leq j \leq J=K-1$. Then the first term in the above EIF can be succinctly written as
\begin{align*}
    v(\bx,a,y)^\top \Gamma \bh^*(\bx) &= v(\bx,a,y)^\top \Gamma L^{-1}(\bx)e_{i-1} \\
    &= v(\bx,a,y)^\top \Gamma \Gamma^{-1} W^{-1}(\bx)(\Gamma^\top)^{-1} e_{i-1} \\
    &= v(\bx,a,y)^\top W^{-1}(\bx)(\Gamma^{-1})^\top e_{i-1} \\
    &= e_{i-1}^\top \Gamma^{-1} W^{-1}(\bx)v(\bx,a,y) \\
    &= \sum_{j=1}^{K-1} \gamma_j^i \frac{I\{a=(k_j,l_j)\}\{y-m_{k_jl_j}(\bx)\}}{\pi((k_j,l_j)|\bx)m_{k_jl_j}(\bx)(1-m_{k_jl_j}(\bx))}.
\end{align*}
where $e_{i-1}$ is a $(K-1)$-dimensional vector with the $(i-1)$-th entry being 1 and all other entries being 0, and we recall that $\gamma_j^i = (\Gamma^{-1})_{i-1,j}$. 
\end{proof}

\noindent\textbf{Case II: $K=2$.} This is a ``sanity check." For $K=2$ there is a single pairwise comparison between players 1 and 2, and $\theta_2(\bx)$ is the negative conditional log odds of player 1 winning over player 2. The propensity score is $\pi((1,2)|\bx) = 1$ for all $\bx$. Hence our system of equations becomes a single equation: $m_{12}(\bx)(1-m_{12}(\bx)) h_2^*(\bx) = 1$. Thus, $h_2^*(\bx) = \{m_{12}(\bx)(1-m_{12}(\bx))\}^{-1}$. Using this in the expression of the EIF, we get
\begin{equation*}
    -\frac{y-m_{12}(\bx)}{m_{12}(\bx)(1-m_{12}(\bx))} + \theta_2(\bx) - \phi_2,
\end{equation*}
which agrees with what we expect if we simply estimated the mean (over covariates) of a transformation of a conditional mean function (of an outcome.)
\\
\\
\noindent\textbf{Case III: $K=3$.} With 3 players, $\calA = \{(1,2),(1,3),(2,3)\}$ and we assume that $P(A=a|\bX=\bx) >0$ for all $a \in \calA$ and $\bx \in \mathcal{X}$. In this case, $L(\bx)$ is a $2\times2$ matrix, and its inverse has an explicit formula. The diagonal elements of this matrix are $g_{12}(\bx) + g_{23}(\bx)$ and $g_{13}(\bx)+g_{23}(\bx)$, and the off-diagonal elements are both $-g_{23}(\bx)$. Determinant of this matrix is $g_{12}(\bx)g_{13}(\bx) + g_{12}(\bx)g_{23}(\bx) + g_{13}(\bx)g_{23}(\bx)$.

For inferring $\phi_2$, we have
\begin{align*}
    h_2^*(\bx) &= \frac{g_{13}(\bx)+g_{23}(\bx)}{g_{12}(\bx)g_{13}(\bx) + g_{12}(\bx)g_{23}(\bx) + g_{13}(\bx)g_{23}(\bx)}; \\
    h_3^*(\bx) &= \frac{g_{23}(\bx)}{g_{12}(\bx)g_{13}(\bx) + g_{12}(\bx)g_{23}(\bx) + g_{13}(\bx)g_{23}(\bx)}.
\end{align*}
Therefore, the EIF of $\phi_2$ is
\begin{align*}
    (\bx,a,y) &\mapsto -\frac{I\{a=(1,2)\}\left\{y-m_{12}(\bx)\right\}\left\{g_{13}(\bx)+g_{23}(\bx)\right\}}{g_{12}(\bx)g_{13}(\bx) + g_{12}(\bx)g_{23}(\bx) + g_{13}(\bx)g_{23}(\bx)} \\
    &\quad - \frac{I\{a=(1,3)\}\left\{y-m_{13}(\bx)\right\}g_{23}(\bx)}{g_{12}(\bx)g_{13}(\bx) + g_{12}(\bx)g_{23}(\bx) + g_{13}(\bx)g_{23}(\bx)} \\
    &\quad + \frac{I\{a=(2,3)\}\left\{y-m_{23}(\bx)\right\}g_{13}(\bx)}{g_{12}(\bx)g_{13}(\bx) + g_{12}(\bx)g_{23}(\bx) + g_{13}(\bx)g_{23}(\bx)} + \theta_2(\bx) - \phi_2,
\end{align*}
where we recall the definition of $g_{kl}$, $g_{kl}(\bx) = \pi((k,l)|\bx)m_{kl}(\bx)(1-m_{kl}(\bx))$, for $1\leq k < l \leq K$; and $g_{lk}(\bx) = g_{kl}(\bx)$, which involves the propensity $\pi((k,l)|\bx)$ and the conditional variance of $Y$.

\section{Proof of all theoretical results}\label{app:proof}

\begin{proof}[Proof of Proposition~\ref{prop: KL projection}]
The log-likelihood of $P_\theta \in \mathcal{P}$ is
\begin{align*}
    \log p_\theta(\bx,a,y) &= \log p_{\bX}(\bx) + \sum_{1 \leq k < l \leq K}I\{a=(k,l)\}\log p_{A|\bX}((k,l)|\bx) \\
    &\quad + \sum_{1 \leq k < l \leq K}I\{a=(k,l)\}\left\{y \log \sigma\left(\theta_k(\bx)-\theta_l(\bx)\right) + (1-y)\log \left(1-\sigma\left(\theta_k(\bx)-\theta_l(\bx)\right)\right)\right\},
\end{align*}
and the log-likelihood of the target distribution $Q$ is
\begin{equation*}
    \log q_\theta(\bx,a,y) = \log q_{\bX}(\bx) + \sum_{1 \leq k < l \leq K}I\{a=(k,l)\}\left\{\log \rho_{kl}(\bx) + y \log q_{kl}(\bx) + (1-y)\log \left(1-q_{kl}(\bx)\right)\right\}.
\end{equation*}
First, as $P_{\bX}$ and $P_{A|\bX}$ are unrestricted under $\mathcal{P}$, to minimize the KL divergence, one need to choose $p_{\bX}(\bx) = q_{\bX}(\bx)$ and $p_{A|\bX}((k,l)|\bx) = \rho_{kl}(\bx)$ for $Q$-almost every $\bx$. We thus focus on minimizing the part of the KL divergence involving $\theta$.

\begin{align*}
    &\quad \EE_Q\left[\sum_{1 \leq k < l \leq K}I\{a=(k,l)\}\left\{y\log \frac{q_{kl}(\bx)}{\sigma\left(\theta_k(\bx)-\theta_l(\bx)\right)} + (1-y) \log \frac{1-q_{kl}(\bx)}{1-\sigma\left(\theta_k(\bx)-\theta_l(\bx)\right)}\right\}\right] \\
    &=\EE_{Q_{\bX}}\left[\sum_{1 \leq k < l \leq K}\rho_{kl}(\bx)\left\{q_{kl}(\bx)\log \frac{q_{kl}(\bx)}{\sigma\left(\theta_k(\bx)-\theta_l(\bx)\right)} + (1-q_{kl}(\bx)) \log \frac{1-q_{kl}(\bx)}{1-\sigma\left(\theta_k(\bx)-\theta_l(\bx)\right)}\right\}\right].
\end{align*}
To minimize the expectation in the second line of the above display, it suffices to minimize the integrand for $Q_{\bX}$-almost every $\bx$. 

Define a function $g_{\bx}$ indexed by $\bx$ such that 
\begin{equation*}
    g_{\bx}(\theta_1,\ldots,\theta_K) = \sum_{1 \leq k < l \leq K}\rho_{kl}(\bx)\left\{q_{kl}(\bx)\log \frac{q_{kl}(\bx)}{\sigma\left(\theta_k-\theta_l\right)} + (1-q_{kl}(\bx)) \log \frac{1-q_{kl}(\bx)}{1-\sigma\left(\theta_k-\theta_l\right)}\right\}.
\end{equation*}
Collecting all terms involving $\theta_k$ and differentiating with respect to $\theta_k$, we get
\begin{align*}
    \frac{\partial g_{\bx}}{\partial \theta_k} &= \sum_{l>k}\rho_{kl}(\bx)\left\{\sigma\left(\theta_k-\theta_l\right) - q_{kl}(\bx) \right\} + \sum_{l < k} \rho_{lk}(\bx)\left\{-\sigma\left(\theta_l-\theta_k\right) + q_{lk}(\bx) \right\} \\
    &= \sum_{l \neq k}\rho_{kl}(\bx)\left\{\sigma\left(\theta_k-\theta_l\right) - q_{kl}(\bx) \right\}.
\end{align*}
Setting all partial derivatives to 0, we get the desired results.
\end{proof}

\begin{proof}[Proof of Proposition~\ref{prop: EIF no shift}]
 Under Assumption~\ref{cond: no covariate shift}, we have $q_{kl}(\bx) = \EE[Y^{kl}|\bX=\bx] = P(Y=1|A=(k,l),\bX=\bx) := m_{kl}(\bx)$ for $1 \leq k < l \leq K$, and let $m_{lk}(\bx) = 1 - m_{kl}(\bx)$. Let $\bU(\btheta(\bx),\bm(\bx)) \in \mathbb{R}^{K-1}$ with the $(k-1)$st component $U_k(\btheta(\bx),\bm(\bx))$ defined as
\begin{equation*}
    U_k(\btheta(\bx),\bm(\bx)) = \sum_{l \neq k} \rho_{kl}(\bx)\left\{\sigma\left(\theta_k(\bx) - \theta_l(\bx)\right) - m_{kl}(\bx)\right\}, \quad \textnormal{for } 2 \leq k \leq K.
\end{equation*}
Proposition~\ref{prop: KL projection} implies that for $P_{\bX}$-almost every $\bx$, $\btheta^*(\bx)$ is implicitly defined as the solution to the equations $\bU(\btheta,\bm(\bx)) = 0$, and $\bphi = \EE_{P_{\bX}}[\btheta^*(\bX)]$.

To start, we notice that $\bphi$ does not depend on the conditional distribution of $A$ given $\bX$. Hence, for the purpose of deriving the efficient influence function, we can treat $P_{A|\bX}$ as known and consider a restricted model. Let $\pi(a|\bx) = P_{A|\bX}(A=a|\bX=\bx)$. The tangent space of the restricted model is the closure of the linear span of $\{g(\bx) + h(\bx,a,y): g,h \in L^2_0(P), \EE_P[h(\bX,A,Y|A=a,\bX=\bx)] = 0 \textnormal{ for all } a \in \mathcal{A}, \bx \in \mathcal{X}\}$. 

Consider a perturbed distribution $P_\epsilon$ with density $$p_\epsilon(\bx,a,y) = p_{\bX}(\bx)(1+\epsilon h(\bx))\pi(a|\bx)p_{Y|(A,\bX)}(y|a,\bx)(1+\epsilon h(y|a,\bx)),$$
whose score with respect to $\epsilon$ is $h(\bx) + h(y|a,\bx)$. Let $m_{kl,\epsilon}(\bx) = P_\epsilon(Y=1|A=(k,l),\bX=\bx)$ be the conditional mean of $Y$ under the perturbed distribution, and let $\btheta^*_\epsilon(\bx)$ be the solution to the equations $\bU(\btheta,\bm_\epsilon(\bx)) = 0$. Then, by chain rule, we get
\begin{align*}
    0 = \frac{\partial \bU(\btheta^*_\epsilon(\bx),\bm_\epsilon(\bx))}{\partial \epsilon} = \frac{\partial \bU(\btheta^*_\epsilon(\bx),\bm_\epsilon(\bx))}{\partial \btheta^*_\epsilon(\bx)}
    \frac{\partial \btheta^*_\epsilon(\bx)}{\partial \epsilon} + 
    \frac{\partial \bU(\btheta^*_\epsilon(\bx),\bm_\epsilon(\bx))}{\partial \bm_\epsilon(\bx)}
    \frac{\partial \bm_\epsilon(\bx)}{\partial \epsilon}.
\end{align*}
Furthermore, for $1 \leq k < l \leq L$,
\begin{equation*}
    \frac{\partial m_{kl,\epsilon}(\bx)}{\partial \epsilon} = \frac{\partial}{\partial \epsilon} \int y p_{Y|(A,\bX)}(y|(k,l),\bx)(1+\epsilon h(y|(k,l),\bx))dy = \int y p_{Y|(A,\bX)}(y|(k,l),\bx)h(y|(k,l),\bx)dy,
\end{equation*}
and $\partial m_{lk,\epsilon}(\bx)/\partial \epsilon = -\partial m_{kl,\epsilon}(\bx)/\partial \epsilon$. This result can be equivalently written as, for all $k \neq l$,
\begin{align*}
    \frac{\partial m_{kl,\epsilon}(\bx)}{\partial \epsilon} &= \int (-1)^{I\{k < l\}+1} y p_{Y|(A,\bX)}(y|A=(k\wedge l, k\vee l),\bX=\bx) h(y|(k\wedge l, k\vee l),\bx)dy \\
    &= \int (-1)^{I\{k < l\}+1} \left\{y - \EE\left[Y|A=(k\wedge l, k\vee l),\bX=\bx\right]\right\} p_{Y|(A,\bX)}(y|A=(k\wedge l, k\vee l),\bX=\bx) h(y|(k\wedge l, k\vee l),\bx)dy \\
    &= \EE_P\left[ (-1)^{I\{k < l\}+1} \frac{I\left\{A = (k\wedge l, k\vee l)\right\}}{\pi((k\wedge l, k\vee l)|\bX)} \left\{Y - \EE\left[Y|A=(k\wedge l, k\vee l),\bX\right]\right\} h(Y|A,\bX) \mid \bX=\bx \right] \\
    &= \EE_P\left[ (-1)^{I\{k < l\}+1} \frac{I\left\{A = (k\wedge l, k\vee l)\right\}}{\pi((k\wedge l, k\vee l)|\bX)} \left\{Y - \EE\left[Y|A=(k\wedge l, k\vee l),\bX\right]\right\} \left\{ h(Y|A,\bX) + h(\bX) \right\} \mid \bX=\bx \right] \\
    &= \EE_P \left[\tau_{kl}(\bX,A,Y) \left\{ h(Y|A,\bX) + h(\bX) \right\} \mid \bX=\bx \right],
\end{align*}
where
\begin{equation*}
    \tau_{kl}(\bx,a,y) =  (-1)^{I\{k < l\}+1} \frac{I\left\{a = (k\wedge l, k\vee l)\right\}}{\pi((k\wedge l, k\vee l)|\bx)} \left\{y - \EE\left[Y|A=(k\wedge l, k\vee l),\bX = \bx\right]\right\}.
\end{equation*}
Now, define functions $\tau: \{0,1\}\times \calA \times \mathcal{X} \rightarrow \mathbb{R}^{(K-1)^2}$ and $\tau_k: \{0,1\}\times \calA \times \mathcal{X} \rightarrow \mathbb{R}^{(K-1)}$ such that $\tau(\bx,a,y) = (\tau_2(\bx,a,y)^\top, \ldots, \tau_{K}(\bx,a,y)^\top)^\top$ and $\tau_k(\bx,a,y) = (\tau_{k1}(\bx,a,y),\ldots,\tau_{k(k-1)}(\bx,a,y),\tau_{k(k+1)}(\bx,a,y),\ldots,\tau_{kK}(\bx,a,y))^\top$. Then, we get
\begin{equation*}
    \frac{\partial \bm_{\epsilon}(\bx)}{\partial \epsilon} = \EE_P\left[\tau(\bX,A,Y) \left\{ h(Y|A,\bX) + h(\bX) \right\} \mid \bX=\bx \right].
\end{equation*}

Given that $\bphi(P_\epsilon) = \int \btheta^*_\epsilon(\bx)p_{\bX}(\bx)(1+\epsilon h(\bx))d\bx$, we have
\begin{align*}
    \frac{\partial \bphi(P_\epsilon)}{\partial \epsilon}\Bigr\rvert_{\epsilon=0} &= \int \btheta^*(\bx)h(\bx)p_{\bX}(\bx)d\bx + \int \frac{\partial \btheta^*_\epsilon(\bx)}{\partial \epsilon}\Bigr\rvert_{\epsilon=0} p_{\bX}(\bx)d\bx \\
    &= \int \btheta^*(\bx)h(\bx)p_{\bX}(\bx)d\bx \\
    &\quad - \int \left\{\frac{\partial \bU(\btheta^*(\bx),\bm(\bx))}{\partial \btheta^*(\bx)}\right\}^{-1}
    \left\{\frac{\partial \bU(\btheta^*(\bx),\bm(\bx))}{\partial \bm(\bx)}\right\}
    \left\{\frac{\partial \bm_\epsilon(\bx)}{\partial \epsilon}\Bigr\rvert_{\epsilon=0} \right\}p_{\bX}(\bx)d\bx \\
    &= \EE_P\left[\btheta^*(\bX)\left\{h(\bX)+h(Y|A,\bX)\right\}\right] \\
    &\quad - \EE_P\left[ \left\{\frac{\partial \bU(\btheta^*(\bX),\bm(\bX))}{\partial \btheta^*(\bX)}\right\}^{-1}
    \left\{\frac{\partial \bU(\btheta^*(\bX),\bm(\bX))}{\partial \bm(\bX)}\right\}
    \left\{\frac{\partial \bm_\epsilon(\bX)}{\partial \epsilon}\Bigr\rvert_{\epsilon=0} \right\}\right] \\
    &= \EE_P\left[\btheta^*(\bX)\left\{h(\bX)+h(Y|A,\bX)\right\}\right] \\
    &\quad - \EE_P\left[ \left\{\frac{\partial \bU(\btheta^*(\bX),\bm(\bX))}{\partial \btheta^*(\bX)}\right\}^{-1}
    \left\{\frac{\partial \bU(\btheta^*(\bX),\bm(\bX))}{\partial \bm(\bX)}\right\}
    \EE_P\left[\tau(\bX,A,Y) \left\{ h(Y|A,\bX) + h(\bX) \right\} \mid \bX \right]
    \right] \\
    &= \EE_P\left[\left(\btheta^*(\bX) - \bphi - \left\{\frac{\partial \bU(\btheta^*(\bX),\bm(\bX))}{\partial \btheta^*(\bX)}\right\}^{-1}
    \left\{\frac{\partial \bU(\btheta^*(\bX),\bm(\bX))}{\partial \bm(\bX)}\right\}
    \tau(\bX,A,Y) \right) \left\{ h(Y|A,\bX) + h(\bX) \right\} \right]
\end{align*}
Therefore, we see that
\begin{equation*}
    D(\bx,a,y) = - \left\{\frac{\partial \bU(\btheta^*(\bx),\bm(\bx))}{\partial \btheta^*(\bx)}\right\}^{-1}
    \left\{\frac{\partial \bU(\btheta^*(\bx),\bm(\bx))}{\partial \bm(\bx)}\right\}
    \tau(\bx,a,y) + \btheta^*(\bx) - \bphi
\end{equation*}
is an influence function of $\bphi$. Furthermore, it is easy to see that it belongs to the tangent space of the restricted model. Therefore, it is the efficient influence function with respect to the restricted model where $P_{A|\bX}$ is known, which coincides with the efficient influence function in the original locally nonparametric model.

We give explicit expressions for the relevant derivative matrices in the EIF. Specifically, $\partial \bU(\btheta^*(\bx),\bm(\bx))/\partial \btheta^*(\bx)$ is a $(K-1)\times (K-1)$ matrix (noting that $\theta^*_1 = 0$,) and the $(k-1)$-th diagonal entry for $k \in \{2,\ldots, K\}$ is 
\begin{equation*}
    \sum_{l \neq k} \rho_{kl}(\bx)\sigma\left(\theta_k(\bx) - \theta_l(\bx)\right)\left\{1-\sigma\left(\theta_k(\bx) - \theta_l(\bx)\right)\right\},
\end{equation*}
$(k-1,l-1)$-th off diagonal entry for $k,l \in \{2,\ldots, K\}, k\neq l$ is
\begin{equation*}
    -\rho_{kl}(\bx)\sigma\left(\theta_k(\bx) - \theta_l(\bx)\right)\left\{1-\sigma\left(\theta_k(\bx) - \theta_l(\bx)\right)\right\}.
\end{equation*}
The matrix $\partial \bU(\btheta^*(\bx),\bm(\bx))/\partial \bm(\bx)$ is a $(K-1)\times (K-1)^2$ matrix with a block diagonal structure. In particular, $\partial \bU(\btheta^*(\bx),\bm(\bx))/\partial \bm(\bx)$ has $K-1$ diagonal blocks with the $(k-1)$-th diagonal block being $(-\rho_{k1}(\bx),\ldots, -\rho_{k(k-1)}(\bx), -\rho_{k(k+1)}(\bx),\ldots, -\rho_{kK}(\bx))$ for $k\in\{2,\ldots,K\}$ and all off diagonal blocks being $\boldsymbol{0}$.

\paragraph{EIF of $\bpsi$.} Now we derive the EIF of $\bpsi$ in a similar fashion. First, let $U(\bpsi,\bm) \in \mathbb{R}^{K-1}$ with the $(k-1)$st component $U_k(\bpsi,\bm)$ defined as 
\begin{equation*}
    U_k(\bpsi,\bm) = \sum_{l \neq k} \rho_{kl}\left\{\sigma\left(\psi_k - \psi_l\right) - m_{kl}\right\},
\end{equation*}
where $m_{kl} = \EE_P[m_{kl}(\bX)]$ and $\bm$ is the concatenation of $\{m_{kl}:k\neq l\}$. By definition, $U(\bpsi_\epsilon, \bm_\epsilon) = \boldsymbol{0}$, and hence
\begin{equation*}
    \frac{\partial U}{\partial \bpsi} \frac{\partial \bpsi_\epsilon}{\partial \epsilon} \Bigr\rvert_{\epsilon = 0} + \frac{\partial U}{\partial \bm} \frac{\partial \bm_\epsilon}{\partial \epsilon} \Bigr\rvert_{\epsilon = 0} = 0,
\end{equation*}
and
\begin{align*}
    \frac{\partial \bm_\epsilon}{\partial \epsilon} &= \frac{\partial}{\partial \epsilon} \int \bm_\epsilon(\bx) p(\bx)(1+\epsilon h(\bx))d\bx \\
    &= \int \frac{\partial \bm_\epsilon(\bx)}{\partial \epsilon}p(\bx)d\bx + \int \bm(\bx)h(\bx)p(\bx)d\bx \\
    &= \EE_P\left[\EE_P\left[\tau(\bX,A,Y) \left\{ h(Y|A,\bX) + h(\bX) \right\} \mid \bX \right]\right] + \EE_P\left[\left\{\bm(\bX) - \bm\right\}h(\bX)\right] \\
    &= \EE_P\left[\left\{\tau(\bX,A,Y) + \bm(\bX) - \bm\right\}\left\{ h(Y|A,\bX) + h(\bX) \right\} \right].
\end{align*}
Combining these results, we get
\begin{equation*}
    \frac{\partial \bpsi_\epsilon}{\partial \epsilon} = \left\{-\frac{\partial U}{\partial \bpsi}\right\}^{-1}\left\{\frac{\partial U}{\partial \bm}\right\}\EE_P\left[\left\{\tau(\bX,A,Y) + \bm(\bX) - \bm\right\}\left\{ h(Y|A,\bX) + h(\bX) \right\} \right],
\end{equation*}
and hence by definition the following function is an influence function of $\bpsi$:
\begin{equation*}
    D_{\bpsi}: (\bx,a,y) \mapsto \left\{-\frac{\partial U}{\partial \bpsi}\right\}^{-1}\left\{\frac{\partial U}{\partial \bm}\right\}\left\{\tau(\bx,a,y) + \bm(\bx) - \bm\right\}.
\end{equation*}
It is easy to verify that $D_{\bpsi}$ lies in the tangent space in the reduced model where the conditional distribution of $A|\bX$ is known, and thus it is the efficient influence function with respect to this reduced model, which coincides with the efficient influence function in the locally nonparametric model. The derivative matrices have the following explicit forms. The matrix $\partial \bU/\partial \bm \in \mathbb{R}^{(K-1)\times (K-1)^2}$ is block diagonal with $K-1$ diagonal blocks and all off diagonal blocks being $\boldsymbol{0}$, and the $(k-1)$-th diagonal block is $(-\rho_{k1},\ldots, -\rho_{k(k-1)}, -\rho_{k(k+1)},\ldots, -\rho_{kK})$ for $k\in\{2,\ldots,K\}$. The matrix $\partial \bU/\partial \bpsi \in \mathbb{R}^{(K-1)\times (K-1)}$ with the $(k-1)$-th diagonal entry for $k \in \{2,\ldots, K\}$ being $\sum_{l \neq k} \rho_{kl}\sigma(\psi_k - \psi_l)(1-\sigma(\psi_k - \psi_l))$ and the $(k-1,l-1)$-th off diagonal entry for $k,l \in \{2,\ldots, K\}, k\neq l$ being $-\rho_{kl}\sigma(\psi_k - \psi_l)(1-\sigma(\psi_k - \psi_l))$.

\end{proof}

\begin{proof}[Proof of Proposition~\ref{prop: EIF with shift}]
Given Proposition~\ref{prop: EIF no shift}, we apply Theorem 2 and Corollary 1 in \cite{li2023efficient} and obtain the EIF in the data fusion setting.
\end{proof}

\begin{proof}[Proof of Proposition~\ref{prop: IF conditional BT}]
First we consider the situation without covariate shift and $P_{\bX} = Q_{\bX}$. We derive the influence function of $\bphi$ relative to a locally nonparametric model with known conditional distribution of $A|\bX$. This influence function remains an influence function under the conditional BT model with known distribution of $A|\bX$ as it is a more restricted model.

Consider the following perturbed distribution again
$$p_\epsilon(\bx,a,y) = p_{\bX}(\bx)(1+\epsilon h(\bx))\pi(a|\bx)p_{Y|(A,\bX)}(y|a,\bx)(1+\epsilon h(y|a,\bx)),$$
whose score with respect to $\epsilon$ is $h(\bx) + h(y|a,\bx)$.
\begin{align*}
    \frac{\partial \bphi(P_\epsilon)}{\partial \epsilon}\Bigr\rvert_{\epsilon=0} &= \frac{\partial}{\partial\epsilon}\Bigr\rvert_{\epsilon=0} \int \Gamma^{-1}\sigma^{-1}(\bm_{\Gamma,\epsilon}(\bx)) p_{\bX}(\bx)(1+\epsilon h(\bx))d\bx \\
    &= \int \left\{ \Gamma^{-1}\sigma^{-1}(\bm_\Gamma(\bx)) - \bphi\right\}h(\bx)p_{\bX}(\bx)d\bx + \int \Gamma^{-1} (\sigma^{-1})^{\prime}(\bm_\Gamma(\bx)) \frac{\partial \bm_{C,\epsilon}(\bx)}{\partial \epsilon}\Bigr\rvert_{\epsilon=0} p_{\bX}(\bx)d\bx,
\end{align*}
where $(\sigma^{-1})^{\prime}(\bm_\Gamma(\bx))$ is a $J\times J$ diagonal matrix with the $j$-th diagonal entry given by $m_{k_jl_j}(\bx)^{-1}(1-m_{k_jl_j}(\bx))^{-1}$. Recall from the proof of Proposition~\ref{prop: EIF no shift}, we have
\begin{equation*}
    \frac{\partial m_{kl,\epsilon}(\bx)}{\partial \epsilon} = \EE_P \left[\tau_{kl}(\bX,A,Y) \left\{ h(Y|A,\bX) + h(\bX) \right\} \mid \bX=\bx \right]
\end{equation*}
for any $k,l$, which implies that
\begin{equation*}
    \frac{\partial m_{k_jl_j,\epsilon}(\bx)}{\partial \epsilon} = \EE_P \left[\tau_{k_jl_j}(\bX,A,Y) \left\{ h(Y|A,\bX) + h(\bX) \right\} \mid \bX=\bx \right]
\end{equation*}
Now, define a function $\tilde\tau: (\bx,a,y) \mapsto \tilde\tau(\bx,a,y)\in\mathbb{R}^{J}$ such that 
$$\tilde\tau(\bx,a,y) = (\tilde\tau_1(\bx,a,y),\ldots, \tilde\tau_J(\bx,a,y)),$$ and 
$$\tilde\tau_j(\bx,a,y) = \tau_{k_jl_j}(\bx,a,y)m_{k_jl_j}(\bx)^{-1}(1-m_{k_jl_j}(\bx))^{-1}.$$ We then have that
\begin{align*}
    \frac{\partial \bphi(P_\epsilon)}{\partial \epsilon}\Bigr\rvert_{\epsilon=0} &= \EE_P\left[\left\{\Gamma^{-1}\sigma^{-1}(\bm_\Gamma(\bX)) - \bphi\right\}\left\{h(X) + h(Y|A,X)\right\}\right] \\
    &\quad + \EE_P\left[\Gamma^{-1}\EE_P \left[\tilde\tau(\bX,A,Y) \left\{ h(Y|A,\bX) + h(\bX) \right\} \mid \bX\right]\right] \\
    &= \EE_P\left[\left\{\Gamma^{-1}\tilde\tau(\bX,A,Y) + \Gamma^{-1}\sigma^{-1}(\bm_\Gamma(\bX)) - \bphi\right\}\left\{ h(Y|A,\bX) + h(\bX) \right\}\right].
\end{align*}
Hence, by definition,
\begin{equation*}
    D_{condBT}: (\bx,a,y) \mapsto \Gamma^{-1}\tilde\tau(\bx,a,y) + \Gamma^{-1}\sigma^{-1}(\bm_\Gamma(\bx)) - \bphi
\end{equation*}
is an influence function of $\bphi$.

In the data fusion setting, we again apply Corollary 1 in \cite{li2023efficient} and obtain an influence function
\begin{equation*}
    D_{condBT,f}: (s,\bx,a,y) \mapsto \frac{I\{s=1\}}{Pr(S=1)}\frac{dQ_{\bX}}{dP_{\bX}}(\bx)\Gamma^{-1}\tilde\tau(\bx,a,y) + \frac{I\{s=0\}}{Pr(S=0)}\left\{\Gamma^{-1}\sigma^{-1}(\bm_\Gamma(\bx)) - \bphi\right\}.
\end{equation*}
\end{proof}

\begin{proof}[Proof of Proposition~\ref{prop: IF psi cond BT}]
Next, we turn to influence functions of $\bpsi$. Note that $\bpsi$ is a function of $q_{kl}$ implicitly defined via \eqref{eq: marginal EE}, where $q_{kl} = \EE_Q[q_{kl}(\bX)] = \EE_P[m_{kl}(\bX)]$ under Assumption~\ref{cond: no covariate shift}. Since $m_{kl}(\bx)$ can now be identified with fewer pairwise comparisons at each $\bx$, this provides a simpler identification of $\bpsi$ as well. The influence function of $\bpsi$ can be obtained via the delta method given the influence function of $\bm = \EE_P[\bm(\bX)]$. Following the proof of Proposition~\ref{prop: EIF no shift}, we have
\begin{align*}
    \frac{\partial \bm_\epsilon}{\partial \epsilon} &= \frac{\partial}{\partial \epsilon} \int \bm_\epsilon(\bx) p(\bx)(1+\epsilon h(\bx))d\bx = \int \frac{\partial \bm_\epsilon(\bx)}{\partial \epsilon}p(\bx)d\bx + \int \bm(\bx)h(\bx)p(\bx)d\bx,
\end{align*}
and for $k = 2,\ldots, K$, at $\epsilon = 0$,
\begin{align*}
    \frac{\partial \bm_\epsilon(\bx)}{\partial \epsilon} &= \frac{\partial \bm_\epsilon(\bx)}{\partial \btheta_\epsilon(\bx)} \frac{\partial \btheta_\epsilon(\bx)}{\partial\epsilon} = \left\{\frac{\partial \bm_\epsilon(\bx)}{\partial \btheta_\epsilon(\bx)}\right\}\left\{\Gamma^{-1}(\sigma^{-1})^\prime(\bm_\Gamma) \frac{\partial \bm_{C,\epsilon}(\bx)}{\partial \epsilon}\right\}.
\end{align*}
Given the definition of $\bm(\bx) \in \mathbb{R}^{(K-1)^2}$ and $m_{kl}(\bx) = \sigma(\theta_k(\bx)-\theta_l(\bx))$, we have that the derivative matrix evaluated at $\epsilon=0$ has the following form
\begin{align*}
    \frac{\partial\bm(\bx)}{\partial\btheta(\bx)} &= \left(\frac{\partial\bm_{2 }(\bx)}{\partial\btheta(\bx)} \ \ldots \ \frac{\partial\bm_{K }(\bx)}{\partial\btheta(\bx)}\right); \\
    \frac{\partial\bm_{k }(\bx)}{\partial\btheta(\bx)} &= \left(\frac{\partial m_{k1 }(\bx)}{\partial\btheta(\bx)} \ \ldots \ \frac{\partial m_{k(k-1) }(\bx)}{\partial\btheta(\bx)} \ \frac{\partial m_{k(k+1) }(\bx)}{\partial\btheta(\bx)} \ \ldots \ \frac{\partial m_{kK }(\bx)}{\partial\btheta(\bx)}\right), \quad 2 \leq k \leq K; \\
    \frac{\partial m_{kl }(\bx)}{\partial\theta_k(\bx)}  &= \sigma(\theta_k(\bx)-\theta_l(\bx))\left\{1 - \sigma(\theta_k(\bx)-\theta_l(\bx))\right\}, \quad 2 \leq k \leq K; \\
    \frac{\partial m_{kl }(\bx)}{\partial\theta_l(\bx)}  &= -\sigma(\theta_k(\bx)-\theta_l(\bx))\left\{1 - \sigma(\theta_k(\bx)-\theta_l(\bx))\right\}, \quad 2 \leq k,l \leq K, l \neq k.
\end{align*}
Therefore, following the derivation of the IF of $\bphi$, we similarly have
\begin{align*}
    \frac{\partial \bm_\epsilon}{\partial \epsilon}\Bigr\rvert_{\epsilon=0} &= \EE_P\left[\bm(\bX)\left\{h(\bX) + h(Y|A,\bX)\right\}\right] \\
    &\quad + \EE_P\left[\frac{\partial\bm(\bX)}{\partial\btheta(\bX)} \Gamma^{-1}(\sigma^{-1})^\prime(\bm_\Gamma) \frac{\partial \bm_{C,\epsilon}(\bX)}{\partial \epsilon}\right] \\
    &= \EE_P\left[\left\{\bm(\bX) - \bm\right\}\left\{h(\bX) + h(Y|A,\bX)\right\}\right] \\
    &\quad + \EE_P\left[\frac{\partial\bm(\bX)}{\partial\btheta(\bX)} \Gamma^{-1}\EE_P \left[\tilde\tau(\bX,A,Y) \left\{ h(Y|A,\bX) + h(\bX) \right\} \mid \bX\right]\right] \\
    &= \EE_P\left[\left\{\frac{\partial\bm(\bX)}{\partial\btheta(\bX)} \Gamma^{-1}\tilde\tau(\bX,A,Y) + \bm(\bX) - \bm\right\}\left\{ h(Y|A,\bX) + h(\bX) \right\}\right],
\end{align*}
and
\begin{align*}
    \frac{\partial \bpsi_\epsilon}{\partial \epsilon} &= \left\{-\frac{\partial U}{\partial \bpsi}\right\}^{-1}\left\{\frac{\partial U}{\partial \bm}\right\}\frac{\partial \bm_\epsilon}{\partial \epsilon} \\
    &=  \left\{-\frac{\partial U}{\partial \bpsi}\right\}^{-1}\left\{\frac{\partial U}{\partial \bm}\right\}\EE_P\left[\left\{\frac{\partial\bm(\bX)}{\partial\btheta(\bX)} \Gamma^{-1}\tilde\tau(\bX,A,Y) + \bm(\bX) - \bm\right\}\left\{ h(Y|A,\bX) + h(\bX) \right\}\right].
\end{align*}
Hence, by definition, the following function is an influence function of $\bpsi$:
\begin{equation*}
    D_{condBT,\bpsi}: (\bx,a,y) \mapsto \left\{-\frac{\partial U}{\partial \bpsi}\right\}^{-1}\left\{\frac{\partial U}{\partial \bm}\right\}\left\{\frac{\partial\bm(\bx)}{\partial\btheta(\bx)} \Gamma^{-1}\tilde\tau(\bx,a,y) + \bm(\bx) - \bm\right\}.
\end{equation*}
Applying Theorem 2 in \cite{li2023efficient} yields an IF in the data fusion setting.

The derivation of the EIFs involves similar projections as in the case of $\bphi$ and is omitted for brevity.
\end{proof}

Lastly, we provide the proofs to Theorems~\ref{thm: dr_phi_shift} and \ref{thm: dr_phi_shift_cond} and omit the others due to their similarity. 

\begin{proof}[Proof of Theorem~\ref{thm: dr_phi_shift}]
For the proposed one-step estimator $\widehat{\bphi}$, we have
\begin{align*}
     \widehat{\bphi}_f - \bphi &= P_{N}D_{\bphi,f} + R(\hat{{\PP}},{\PP}) + (\hat{{\PP}} - {\PP})[D_{\bphi,f}(\hat{{\PP}}) -D_{\bphi,f}({\PP}) ],
\end{align*}
where the last term is of second-order when cross-fitting is used for the estimation of nuisance functions. Hence it suffices to examine the remainder term: 
\begin{align*}
 R(\hat{{\PP}},{\PP})
    & = \bPhi(\hat{\PP}) - \bPhi(\PP) + E_{\PP}[D_{\bphi,f}(\hat{\PP})]\\
    & = E_{\PP}\left[- \frac{S}{\widehat{Pr}(S=1)}\hat{w}(\bX) \widehat{\Lambda}(\bX)
    \widehat{\tau}(\bX,A,Y)+ \frac{(1-s)}{\widehat{Pr}(S=0)}\left\{\widehat{\btheta}(\bX) - \bphi\right\} \right]\\
    & = E_{P}\left[- \hat{w}(\bX) \widehat{\Lambda}(\bX)
    \widehat{\tau}(\bX,A,Y)+ w(\bX)\left\{\widehat{\btheta}(\bX) - {\btheta}^*(\bX)\right\} \right]\\
    & = E_{P}\bigg[- \{\hat{w}(\bX) - w(\bX)\} \widehat{\Lambda}(\bX)
    \widehat{\tau}(\bX,A,Y) -  w(\bX)\widehat{\Lambda}(\bX)
    \widehat{\tau}(\bX,A,Y) \\
    & \hspace{3.5em}+ w(\bX)\left\{\widehat{\btheta}(\bX)  - {\btheta}^*(\bX)\right\} \bigg]
    &\intertext{Using Taylor expansion, we have $\widehat{\btheta}(\bx)  - {\btheta}^*(\bx) = \Lambda(\bx) \{\widehat{\bm}(\bx) -{\bm}(\bx) \} + O_{P}(\lVert\widehat{\bm}(\bx) -{\bm}(\bx) \rVert^2 )$. Plugging it back into the above, we have }
    & = E_{P}\bigg[- \{\hat{w}(\bX) - w(\bX)\} \widehat{\Lambda}(\bX)
    \widehat{\tau}(\bX,A,Y) -  w(\bX)\widehat{\Lambda}(\bX)
    \widehat{\tau}(\bX,A,Y) \\
    & \hspace{3.5em}+ w(\bX)\Lambda(\bX)\left\{\widehat{\bm}(\bX)  - {\bm}(\bX)\right\} + O_P(\lVert\widehat{\bm}(\bx) -{\bm}(\bx) \rVert^2)\bigg].
\end{align*}
For ease of notation, let $\tau^\dagger$ be the defined in the same way as $\widehat{\tau}$ except that $y$ is replaced by $\bm(\bx)$. The important note is that inside  $\tau^\dagger$, there is the term $\bm(\bx) - \widehat{\bm}(\bx)$. Continuing the above, we have
\begin{align*}
     R(\hat{{\PP}},{\PP})
    & =  E_{P}\bigg[- \{\hat{w}(\bX) - w(\bX)\} \widehat{\Lambda}(\bX)
    \tau^\dagger(\bX,A,Y) -  w(\bX)\widehat{\Lambda}(\bX)
    \tau^\dagger(\bX,A,Y) \\
    & \hspace{3.5em}+ w(\bX)\Lambda(\bX)\left\{\widehat{\bm}(\bX)  - {\bm}(\bX)\right\} +  O_P(\lVert\widehat{\bm}(\bx) -{\bm}(\bx) \rVert^2) \bigg]\\
    & \intertext{By Cauchy-Schwarz inequality, the above is bounded by (up to a multiplicative factor),}
    & \leq \big( \lVert \hat{w}(\bx) - w(\bx) \rVert \lVert \widehat{\bm}(\bx) - \bm(\bx) \rVert+  \lVert \widehat{\pi}(\bx) - \pi(\bx)\rVert \lVert \widehat{\bm}(\bx) - \bm(\bx)\rVert\\
    & \quad   + \lVert \widehat{\Lambda}(\bx) - \Lambda(\bx)\rVert  \lVert \widehat{\bm}(\bx) - \bm(\bx)\rVert+  \lVert \widehat{\bm}(\bx) - \bm(\bx)\rVert^2.
\end{align*}
\end{proof}

\begin{proof}[Proof to Theorem~\ref{thm: dr_phi_shift_cond}]
Similar to above, we examine the reminder term:
\begin{align*}
 R(\hat{{P}},{P})
    & = \bPhi_f(\hat{\PP}) - \bPhi_f(\PP) + E_{\PP}[D^*_{cond,f}(\hat{\PP})]\\  
    &  = E_{\PP} \left[\frac{S}{\widehat{Pr}(S=1)} \hat{w}(\bx)\left(\Gamma_*^\top \widehat{W}_*(\bx) \Gamma_*\right)^{-1}\Gamma_*^\top \hat{v}(\bx,a,y) + \frac{1-S}{\widehat{Pr}(S=0)}(\widehat{\btheta}^*(\bx) - \bphi )\right] \\
    & = E_{P} \left[\hat{w}(\bx)\Gamma_*^{-1} {\widehat{W}_*}^{-1}(\bx) \hat{v}(\bx,a,y) + w(\bx)(\widehat{\btheta}^*(\bx) - \bphi) \right].
\end{align*}
When the conditional BT model holds, we note that $\widehat{\btheta}^*(\bx) = \Gamma_*^{-1}\sigma^{-1}(\widehat{\bm}_{\Gamma_*}(\bx))$ where $\bm_{\Gamma_*} = (m_{k_{*1}l_{*1}}(\bx),\ldots,m_{k_{*K(K-1)/2}l_{*K(K-1)/2}}(\bx))$. Continuing the above,
\begin{align*}
     & = E_{P} \left[\hat{w}(\bx)\Gamma_*^{-1} {\widehat{W}_*}^{-1}(\bx) \hat{v}(\bx,a,y) + w(\bx)(\Gamma_*^{-1}\sigma^{-1}(\widehat{\bm}_{\Gamma_*}(\bx)) - \Gamma_*^{-1}\sigma^{-1}(\bm_{\Gamma_*}(\bx))) \right]\\
     & = E_{P} \left[\hat{w}(\bx)\Gamma_*^{-1} \widehat{\tilde{\tau}}_{*}(\bx,a,y) + w(\bx)(\Gamma_*^{-1}\sigma^{-1}(\widehat{\bm}_{\Gamma_*}(\bx)) - \Gamma_*^{-1}\sigma^{-1}(\bm_{\Gamma_*}(\bx))) \right],
\end{align*}
where $\widehat{\tilde\tau}_*(\bx,a,y) = (\widehat{\tilde\tau}_{*1}(\bx,a,y),\ldots, \widehat{\tilde\tau}_{*K(K-1)/2}(\bx,a,y)) \in \mathbb{R}^{K(K-1)/2}$ and, with $\tau$ defined in \eqref{eq: definition of tau}, 
$$\widehat{\tilde\tau}_{*j}(\bx,a,y) = \tau_{k_{*j}l_{*j}}(\bx,a,y)m_{k_{*j}l_{*j}}(\bx)^{-1}(1-m_{k_{*j}l_{*j}}(\bx))^{-1}.$$
Noting that $\Gamma_*$ is fixed, we have
\begin{align*}
    &\Gamma_*^{-1}\sigma^{-1}(\widehat{\bm}_{\Gamma_*}(\bx)) - \Gamma_*^{-1}\sigma^{-1}({\bm}_{\Gamma_*}(\bx)) \\
    & = (\sigma^{-1})^{\prime}(\bm_{\Gamma_*}(\bx)) \left\{\widehat{\bm}_{\Gamma_*}(\bx) - {\bm}_{\Gamma_*}(\bx)\right\} + O_P \left( \left\{\widehat{\bm}_{\Gamma_*}(\bx) - {\bm}_{\Gamma_*}(\bx)\right\}^2\right)\\
    & = \bm_{\Gamma_*}(\bx)^{-1}(1-\bm_{\Gamma_*}(\bx))^{-1} \left\{\widehat{\bm}_{\Gamma_*}(\bx) - {\bm}_{\Gamma_*}(\bx)\right\} + O_P\left( \left\{\widehat{\bm}_{\Gamma_*}(\bx) - {\bm}_{\Gamma_*}(\bx)\right\}^2\right).
\end{align*}
Plugging this into the above, we have: 
\begin{align*}
    R(\hat{{P}},{P})& = E_{P} \big[\{\hat{w}(\bx) - w(\bx)\}\Gamma_*^{-1}\widehat{\tilde\tau}_*(\bx,a,y)  + w(\bx)\Gamma_*^{-1}\widehat{\tilde\tau}_*(\bx,a,y)\\
    & \quad + w(\bx)\bm_{\Gamma_*}(\bx)^{-1}(1-\bm_{\Gamma_*}(\bx))^{-1} \left\{\widehat{\bm}_{\Gamma_*}(\bx) - {\bm}_{\Gamma_*}(\bx)\right\} + O_P\left( \left\{\widehat{\bm}_{\Gamma_*}(\bx) - {\bm}_{\Gamma_*}(\bx)\right\}^2\right)\big]
    & \intertext{Noting that there is a $\{\bm_{\Gamma_*}(\bx) - \widehat{\bm}_{\Gamma_*}(\bx)\}\widehat{\bm}_{\Gamma_*}(\bx)^{-1}(1-\widehat{\bm}_{\Gamma_*}(\bx))^{-1}$ term inside $\widehat{\tilde\tau}_*$, and using Cauchy-Schwarz inequality again, this term is bounded by (up to a multiplicative factor)}
    & \leq \lVert \widehat{\bm}_{\Gamma_*}(\bx) - {\bm}_{\Gamma_*}(\bx) \rVert \lVert \hat{\pi}(\bx) - \pi(\bx)\rVert + \lVert  \left\{\widehat{\bm}_{\Gamma_*}(\bx) - {\bm}_{\Gamma_*}(\bx)\right\}^2 \rVert\\
    & \quad \lVert \widehat{\bm}_{\Gamma_*}(\bx) - {\bm}_{\Gamma_*}(\bx) \rVert \lVert \hat{w}(\bx) - w(\bx)\rVert\\
    & =  \lVert \widehat{\btheta}^*(\bx) - \btheta(\bx) \rVert \lVert \hat{\pi}(\bx) - \pi(\bx)\rVert + \lVert  \{\widehat{\btheta}^*(\bx) - \btheta(\bx)\}^2 \rVert\\
    & \quad \lVert \widehat{\btheta}^*(\bx) - \btheta(\bx) \rVert \lVert \hat{w}(\bx) - w(\bx)\rVert,
\end{align*}
where that last line is true since $\widehat{\btheta}^*(\bx) = \Gamma_*^{-1}\sigma^{-1}(\widehat{\bm}_{\Gamma_*}(\bx))$ and $\sigma^{-1}$ is smooth.
\end{proof}

\section{Theoretical guarantees for other proposed estimators}\label{app:other estimator}

We now present the theoretical guarantees for other proposed estimators introduced in Section~\ref{sec: nonparam estimation}. We omit the proofs as they closely resemble the preceding one. 
\begin{theorem}[\textit{Efficient 
estimation of $\bphi$} ] \label{thm: dr_phi_no_shift}
    Suppose the conditional BT model is misspecified, and nuisance functions  $\widehat{\pi}$ and $\widehat{\bm}$ were estimated via cross-fitting. Under Assumption~\ref{cond: no covariate shift} , we have
    \begin{align*}
        \widehat{\bphi} - \bphi &= P_{N}D_{\bphi} + O_{{P}}\big(  \lVert \widehat{\bm}(\bx) - \bm(\bx) \rVert^2\\
    & \quad  +  \lVert \widehat{\pi}(\bx) - \pi(\bx)\rVert \lVert \widehat{\bm}(\bx) - \bm(\bx)\rVert + \lVert \widehat{\Lambda}(\bx) - \Lambda(\bx)\rVert  \lVert \widehat{\bm}(\bx) - \bm(\bx)\rVert  \big).
    \end{align*} 
    Moreover, if the nuisance functions were estimated such that the sum of three product terms above is $o_{P}(1/\sqrt{N})$, then the proposed estimator $\widehat{\bphi}$ is consistent, asymptotically normal and achieve the semiparametric efficiency bound. That is,
    $$\sqrt{N}(\widehat{\bphi} - \bphi) \rightarrow_d \mathrm{N}(\mathbf{0}, \mathrm{cov}_{{P}}(D_{\bphi})).$$
\end{theorem}

\begin{theorem}[\textit{Efficient and doubly robust 
estimation of $\bpsi$} ] \label{thm: dr_psi_no_shift}
    Suppose the conditional BT model is misspecified, and nuisance functions  $\widehat{\pi}$ and $\widehat{\bm}$ were estimated via cross-fitting. Under Assumption~\ref{cond: no covariate shift} , we have
    \begin{align*}
        \widehat{\bpsi} - \bpsi &= P_{N}D_{\bpsi} + O_{{P}}\big(  \lVert \widehat{\pi}(\bx) - \pi(\bx)\rVert \lVert \widehat{\bm}(\bx) - \bm(\bx)\rVert\big).
    \end{align*} 
    Moreover, if the nuisance functions were estimated such that the product term above is $o_{P}(1/\sqrt{N})$, then the proposed estimator $\widehat{\bpsi}$ is consistent, asymptotically normal and achieve the semiparametric efficiency bound. That is,
    $$\sqrt{N}(\widehat{\bpsi} - \bpsi) \rightarrow_d \mathrm{N}(\mathbf{0}, \mathrm{cov}_{{P}}(D_{\bpsi})).$$
\end{theorem}

We now present the theoretical results for the proposed estimators constructed by influence functions proposed in Proposition~\ref{prop: IF conditional BT}  in Section~\ref{sec: cond BT estimation}. 

\begin{theorem}[\textit{Estimation of $\bphi$ when conditional BT holds} ] \label{thm: dr_phi_no_shift_cond}
    Suppose the conditional BT model is correctly specified, and nuisance functions $\widehat{\pi}$ and $\widehat{\bm}$ were estimated via cross-fitting. Under Assumptions~\ref{cond: no shift cond} and \ref{cond: gamma full rank}, we have
    \begin{align*}
        \widehat{\bphi}_{cond} - \bphi &= P_{N}D_{cond} + O_{{P}}( \lVert \widehat{\pi}(\bx) - \pi(\bx)\rVert \lVert \widehat{\bm}(\bx) - \bm(\bx)\rVert + \lVert \widehat{\bm}(\bx) - \bm(\bx)\rVert^2  ).
    \end{align*} 
    Moreover, if the nuisance functions were estimated such that the sum of the two product terms above is $o_{P}(1/\sqrt{N})$, then the proposed estimator $\widehat{\bphi}_{cond}$ is consistent and asymptotically normal. That is,
    $$\sqrt{N}(\widehat{\bphi}_{cond}- \bphi) \rightarrow_d \mathrm{N}(\mathbf{0}, \mathrm{cov}_{{P}}(D_{cond})).$$
\end{theorem}

\begin{theorem}[\textit{Estimation of $\bpsi$ when conditional BT holds} ] \label{thm: dr_psi_no_shift_cond_naive}
    Suppose the conditional BT model is correctly specified, and nuisance functions $\widehat{\pi}$ and $\widehat{\bm}$ were estimated via cross-fitting. Under Assumptions~\ref{cond: no shift cond} and \ref{cond: gamma full rank}, we have
    \begin{align*}
        \widehat{\bpsi}_{cond} - \bpsi &= P_{N}D_{cond, \bpsi} + O_{{P}}( \lVert \widehat{\pi}(\bx) - \pi(\bx)\rVert \lVert \widehat{\bm}(\bx) - \bm(\bx)\rVert  ).
    \end{align*} 
    Moreover, if the nuisance functions were estimated such that the sum of the two product terms above is $o_{P}(1/\sqrt{N})$, then the proposed estimator $\widehat{\bpsi}_{cond}$ is consistent and asymptotically normal. That is,
    $$\sqrt{N}(\widehat{\bpsi}_{cond}- \bpsi) \rightarrow_d \mathrm{N}(\mathbf{0}, \mathrm{cov}_{{P}}(D_{cond,\bpsi})).$$
\end{theorem}

\begin{theorem}[\textit{Estimation of $\bpsi$ under covariate shifts when conditional BT holds}] \label{thm: dr_psi_shift_cond_naive}
    Suppose the conditional BT model is correctly specified, and nuisance functions $\hat{w}$, $\widehat{\pi}$ and $\widehat{\bm}$ were estimated via cross-fitting. Under Assumptions~\ref{cond: no shift cond}(b)-(c), \ref{cond: with shift cond} and~\ref{cond: gamma full rank}, we have
    \begin{align*}
        \widehat{\bpsi}^*_{cond,f} - \bpsi &= P_{N}D^*_{cond,\bpsi,f} + O_{{\PP}}( \lVert \widehat{\pi}(\bx) - \pi(\bx)\rVert \lVert \widehat{\bm}(\bx) - \bm(\bx)\rVert\\
        & \quad +\lVert \widehat{w}(\bx) - w(\bx)\rVert \lVert \widehat{\bm}(\bx) - \bm(\bx)\rVert  ).
    \end{align*} 
    Moreover, if the nuisance functions were estimated such that the sum of the two product terms above is $o_{\PP}(1/\sqrt{N})$, then the proposed estimator $\widehat{\bpsi}^*_{cond,f}$ is consistent and asymptotically normal, and achieves the semiparametric efficiency bound. That is,
    $$\sqrt{N}(\widehat{\bpsi}^*_{cond,f}- \bpsi) \rightarrow_d \mathrm{N}(\mathbf{0}, \mathrm{cov}_{{\PP}}(D^*_{cond,\bpsi,f})).$$
\end{theorem}

\section{Identification and estimation when the density ratio has a known form}\label{app:known density ratio}

We consider an alternative identification strategy in the presence of covariate shift where we assume that the density ratio $dQ_{\bX}/dP_{\bX} = w$ is known up to a scaling factor. In this case, $\bphi$ and $\bpsi$ are still identifiable from the observed data distribution $P$. Although realistically we may not know how to shift the observed contextual distribution for every future task, this approach can be useful when exploring possible covariate distributions by upweighting or downweighting certain observed contexts.

\begin{proposition}[EIF of $\bphi$ under known covariate shift]\label{prop: EIF with known shift}
The EIF with known (un-normalized) density ratio is 
\begin{equation*}
    D_w(\bx,a,y) = \frac{w(\bx)}{\EE_P[w(\bX)]}\left\{- \Lambda(\btheta^*(\bx),\bm(\bx))
    \tau(\bx,a,y)
    + \btheta^*(\bx) - \bphi\right\},
\end{equation*} 
where $dQ_{\bX}(\bx)/dP_{\bX}(\bx) = w(\bx)/\EE_P[w(\bX)]$ and $w(\bx)$ is known.
\end{proposition}

Given i.i.d. sample $\{(\bX_i, A_i, Y_i)\}_{i=1}^n$, we can construct a one-step estimator of $\bphi$ based on the EIF above.

\begin{proof}[Proof of Proposition~\ref{prop: EIF with known shift}]
Let $q_{\bX}(\bx) = w(\bx)p_{\bX}(\bx)/\int w(\bx)p_{\bX}(\bx)d\bx$ with $w(\bx)$ known. 
\begin{equation*}
    \bphi = \EE_{Q_X}[\btheta^*(\bX)] = \int \btheta^*(\bx) q_{\bX}(\bx) d\bx = \int \btheta^*(\bx) \frac{w(\bx)p_{\bX}(\bx)}{\int w(\tilde\bx)p_{\bX}(\tilde\bx)d\tilde\bx} d\bx = \frac{\EE_{P_{\bX}}[w(\bX)\btheta^*(\bX)]}{\EE_{P_{\bX}}[w(\bX)]}.
\end{equation*}
To derive the EIF of $\bphi$, it suffices to derive the EIFs of the numerator and denominator separately. The EIF of the denominator is $(\bx,a,y) \mapsto w(\bx) - \EE_{P_{\bX}}[w(\bX)]$. For the numerator, we consider the same perturbed distribution $P_\epsilon$ as in the proof of Proposition~\ref{prop: EIF no shift}
\begin{align*}
    \frac{\partial \EE_{P_\epsilon}[w(\bX)\btheta^*_\epsilon(\bX)]}{\partial \epsilon}\Bigr\rvert_{\epsilon=0} &= \int w(\bx)\btheta^*(\bx)h(\bx)p_{\bX}(\bx)d\bx + \int w(\bx)\frac{\partial \btheta^*_\epsilon(\bx)}{\partial \epsilon}\Bigr\rvert_{\epsilon=0} p_{\bX}(\bx)d\bx.
\end{align*}
Following the same derivation as in proving Proposition~\ref{prop: EIF no shift} except we now multiple the integrand by $w(\bx)$ and applying the delta method, we get the EIF of $\bphi$ 
\begin{multline*}
    D_w: (\bx,a,y) \mapsto \frac{1}{\EE_P[w(\bX)]}\Bigg[- \left\{\frac{\partial \bU(\btheta^*(\bx),\bm(\bx))}{\partial \btheta^*(\bx)}\right\}^{-1}
    \left\{\frac{\partial \bU(\btheta^*(\bx),\bm(\bx))}{\partial \bm(\bx)}\right\}
    w(\bx)\tau(\bx,a,y) \\
    + w(\bx)\btheta^*(\bx) - \bphi w(\bx)\Bigg].
\end{multline*}
\end{proof}

\section{Additional simulation results}\label{app:sim}

\begin{figure}
    \centering
    \begin{subfigure}[t]{0.49\textwidth}
        \centering
        \includegraphics[width=1.0\textwidth]{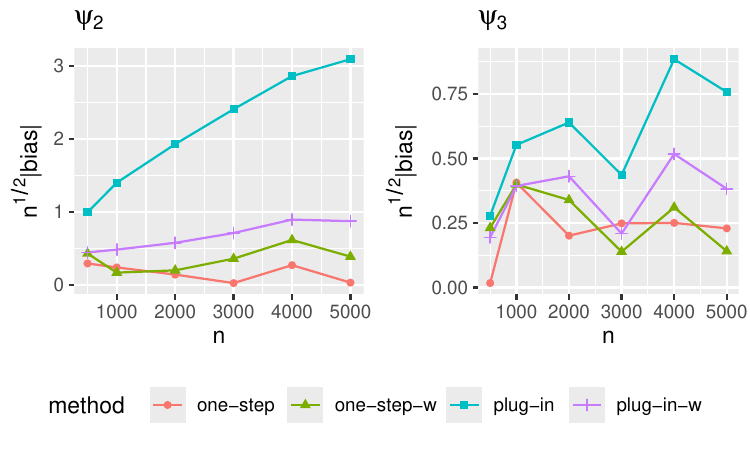}
        \caption{scaled bias of estimators}
    \end{subfigure}%
    ~ 
    \begin{subfigure}[t]{0.49\textwidth}
        \centering
        \includegraphics[width=1.0\textwidth]{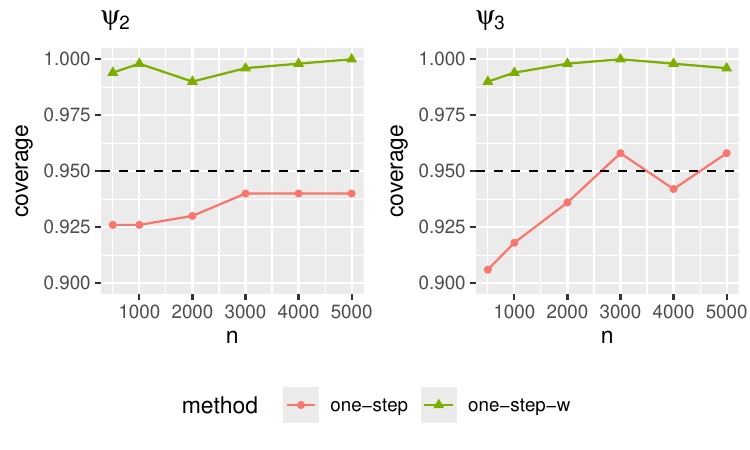}
        \caption{coverage of Wald confidence intervals}
    \end{subfigure}
    \caption{Estimation and inference of $\bpsi$: scaled bias of plug-in and one-step estimators (panel a) and coverage of Wald CI associated with one-step estimators (panel b) under varying sample sizes. ``-w" indicates methods with working parametric models. Results are based on 500 simulation replications.}
    \label{fig: nonparam psi}
\end{figure}

\begin{figure}
    \centering
    \includegraphics[width=0.7\textwidth,height=0.35\textwidth]{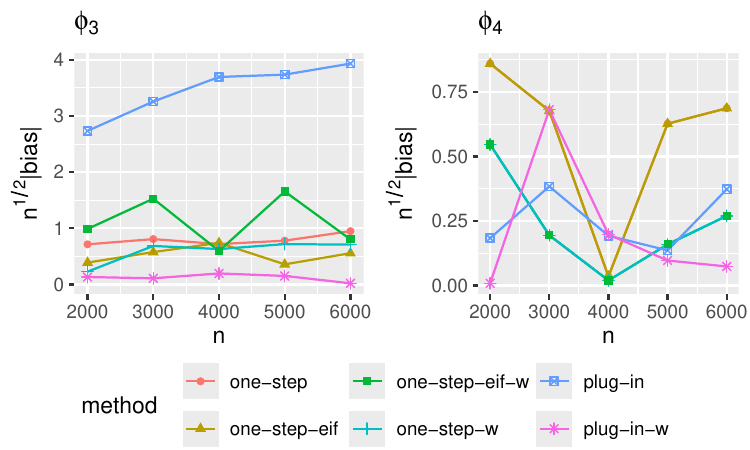}
    \includegraphics[width=0.7\textwidth,height=0.35\textwidth]{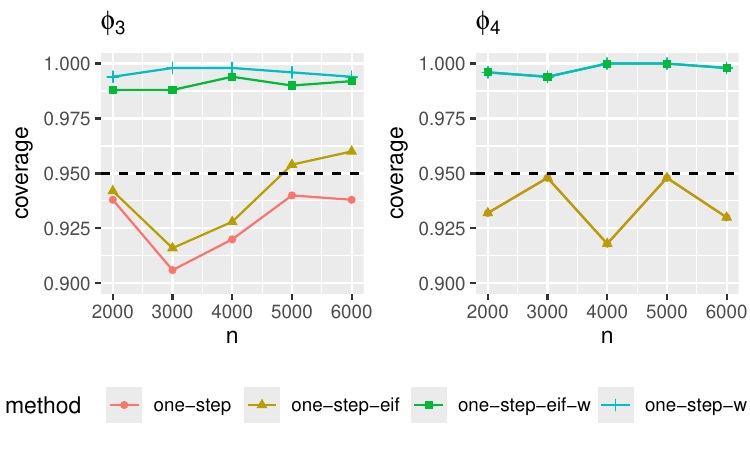}
    \includegraphics[width=0.7\textwidth,height=0.35\textwidth]{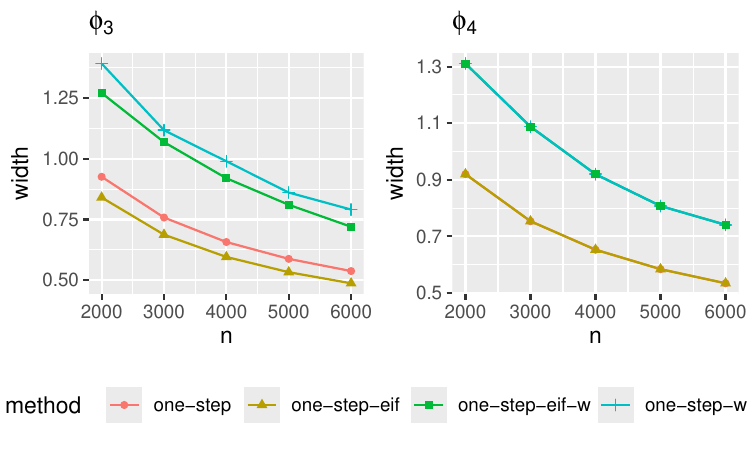}
    \caption{Estimation and inference of $\bphi_3$ and $\bphi_4$: scaled bias of plug-in and one-step estimators (upper panel), coverage (middle panel) and average width (bottom panel) of Wald CI associated with one-step estimators under varying sample sizes. ``-w" indicates methods with working parametric models. Results are based on 500 simulation replications. Data generating distribution belongs to the conditional BT model.}
    \label{fig: cond phi34}
\end{figure}

\begin{figure}
    \centering
    \includegraphics[width=0.7\textwidth,height=0.35\textwidth]{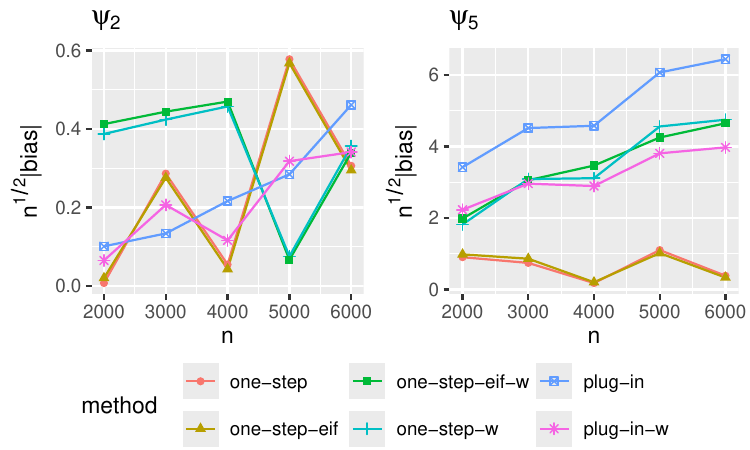}
    \includegraphics[width=0.7\textwidth,height=0.35\textwidth]{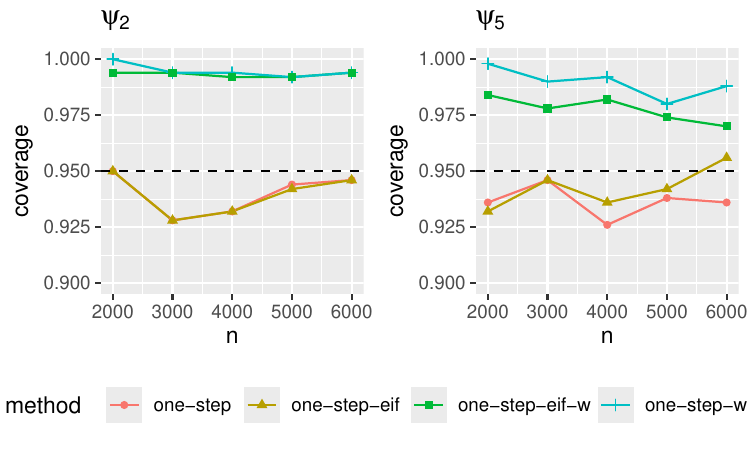}
    \includegraphics[width=0.7\textwidth,height=0.35\textwidth]{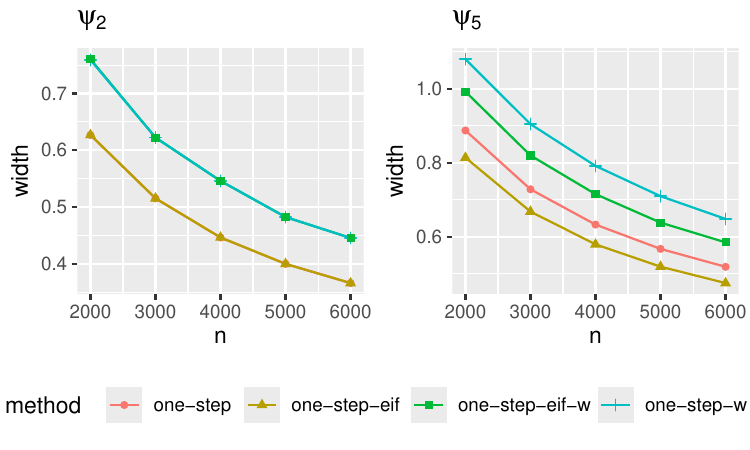}
    \caption{Estimation and inference of $\bpsi_2$ and $\bpsi_5$: scaled bias of plug-in and one-step estimators (upper panel), coverage (middle panel) and average width (bottom panel) of Wald CI associated with one-step estimators under varying sample sizes. ``-w" indicates methods with working parametric models. Results are based on 500 simulation replications. Data generating distribution belongs to the conditional BT model.}
    \label{fig: cond psi25}
\end{figure}

\begin{figure}
    \centering
    \includegraphics[width=0.7\textwidth,height=0.35\textwidth]{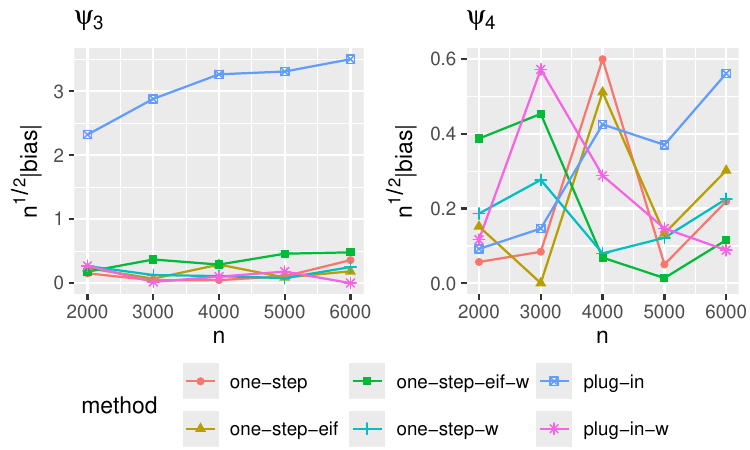}
    \includegraphics[width=0.7\textwidth,height=0.35\textwidth]{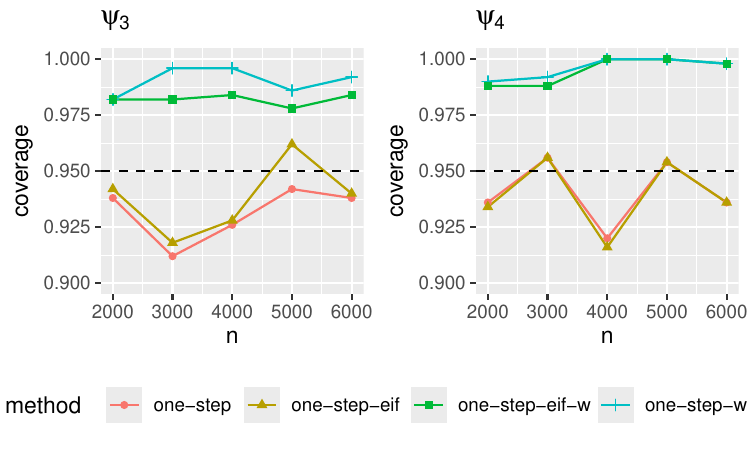}
    \includegraphics[width=0.7\textwidth,height=0.35\textwidth]{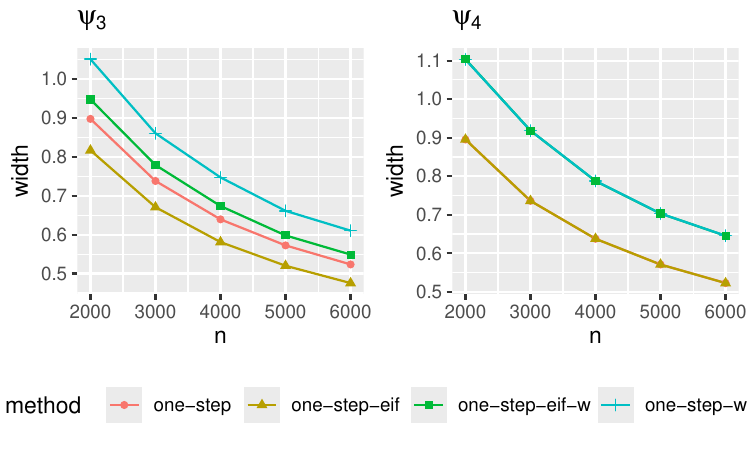}
    \caption{Estimation and inference of $\bpsi_3$ and $\bpsi_4$: scaled bias of plug-in and one-step estimators (upper panel), coverage (middle panel) and average width (bottom panel) of Wald CI associated with one-step estimators under varying sample sizes. ``-w" indicates methods with working parametric models. Results are based on 500 simulation replications. Data generating distribution belongs to the conditional BT model.}
    \label{fig: cond psi34}
\end{figure}

First, we present the results in Figure~\ref{fig: nonparam psi} for estimation of and inference about $\bpsi$ in Setting I where the true data generating distribution does not belong to the marginal BT model nor the conditional BT model. The plug-in estimators and one-step estimators are described in Section~\ref{sec: nonparam estimation}. Next, we turn to Setting II where the true data generating distribution belongs to the conditional BT model. In Figure~\ref{fig: cond phi34} we present the results for inferring $\bphi_3$ and $\bphi_4$, and in Figures~\ref{fig: cond psi25} and \ref{fig: cond psi34} we present the analagous results for $\bpsi$. We observe similar patterns as those described in Section~\ref{subsec: sim}.

\section{Additional data illustration results}\label{app:data}
\begin{table}[H]
    \centering
    \renewcommand{\arraystretch}{1.3}
    \caption{Sample multi-turn questions in MT-bench \cite{zheng2023judging}.}\label{tab:sample_bench}
    \begin{tabular}{|l|l|p{10cm}|} 
        \hline
        \textbf{Category} & & \textbf{Sample Questions} \\
        \hline
        \multirow{2}{*}{Writing} 
        & 1st Turn & Craft an intriguing opening paragraph for a fictional short story. The story should involve a character who wakes up one morning to find that they can time travel. \\
        \cline{2-3}
        & 2nd Turn & Summarize the story with three bullet points using only nouns and adjectives, without verbs. \\
        \hline
        \multirow{2}{*}{Role play} 
        & 1st Turn & Embody the persona of Tony Stark from “Iron Man” throughout this conversation. Bypass the introduction “As Stark”. Our first question is: “What’s your favorite part about being Iron Man?\\
        \cline{2-3}
        & 2nd Turn & What do you think about GPT-4 as a replacement of your JAVIS? \\
        \hline
         \multirow{2}{*}{Reasoning} 
         & 1st Turn & Which word does not belong with the others? tyre, steering wheel, car, engine\\
        \cline{2-3}
        & 2nd Turn & Could you replace it with a word that belongs with the others? \\
         \hline
        \multirow{2}{*}{Extraction} 
        & 1st Turn & Identify the countries, their capitals, and the languages spoken in the following sentences. Output in JSON format... \\
        \cline{2-3}
        & 2nd Turn & Come up with 3 similar examples in the YAML format. \\
         \hline
        \multirow{2}{*}{Mathematics} 
        & 1st Turn & $x+y = 4z$, $x*y = 4z^2$, express $x-y$ in $z$.\\
        \cline{2-3}
        & 2nd Turn & Express $z-x$ in $y$. \\
         \hline
        \multirow{2}{*}{Coding} 
        & 1st Turn & Write a simple website in HTML. When a user clicks the button, it shows a random joke from a list of 4 jokes.\\
        \cline{2-3}
        & 2nd Turn & How to use CSS to change the color of jokes to red? \\
         \hline
        \multirow{2}{*}{STEM} 
        & 1st Turn & What is the central dogma of molecular biology? What processes are involved? Who named this?\\
        \cline{2-3}
        & 2nd Turn & Identify and fix one incorrect fact in your previous response.\\
         \hline
        \multirow{2}{*}{Humanities} 
        & 1st Turn & How do the stages of life shape our understanding of time and mortality?\\
        \cline{2-3}
        & 2nd Turn & Write an allegorical poem that illustrates the above. \\
        \hline
    \end{tabular}
\end{table}

\begin{table}[H]
    \centering
    \caption{Sample multi-turn questions in Chatbot arena crowd-source data \cite{zheng2023judging}.} \label{tab:sample_arena}
    \begin{tabular}{|l|p{10cm}|}
        \hline
        \textbf{Category} & \textbf{Sample Questions} \\
        \hline
        \multirow{1}{*}{Writing} 
        & Write a sci-fi story between Fox McCloud and Wolf O'Donnell in style of Japanese light novel. \\
        \hline
        \multirow{1}{*}{Role play} 
        & You are a german citizen interested in weathering a possible economic downturn in Europe... How do you rank this opportunity against other options?\\
        \hline
         \multirow{1}{*}{Reasoning} 
         & If 1 is 2, 2 is 3, and 5 is 6, does this mean that 4 is necessarily 5\\
         \hline
        \multirow{1}{*}{Extraction} 
        &  Context - The Company does not include intercompany transfers between segments for management reporting purposes. Note 2 – Revenue Net sales disaggregated by significant products and services for 2022, 2021 and 2020 were as follows (in millions): iPhone (1) Mac (1) iPad (1) Wearables, Home and Accessories... Question - what is the sales of iphone in 2021 and 2022 and compare it to total revenue ?
\\
         \hline
        \multirow{1}{*}{Mathematics} 
        & Which is bigger 1534 or 981 ?\\
         \hline
        \multirow{1}{*}{Coding} 
        & Write code for ffmpeg to convert mp4 video to mkv using the h265 codec and crf setting.\\
         \hline
        \multirow{1}{*}{STEM} 
        & what are few shot learning? \\
         \hline
        \multirow{1}{*}{Humanities} 
        & Enumerate the nine realms of the norse mythology. Only enumerate them, do not explain them.\\
        \hline
    \end{tabular}
\end{table}

\begin{figure}[H]
    \centering
    \includegraphics[width=\linewidth]{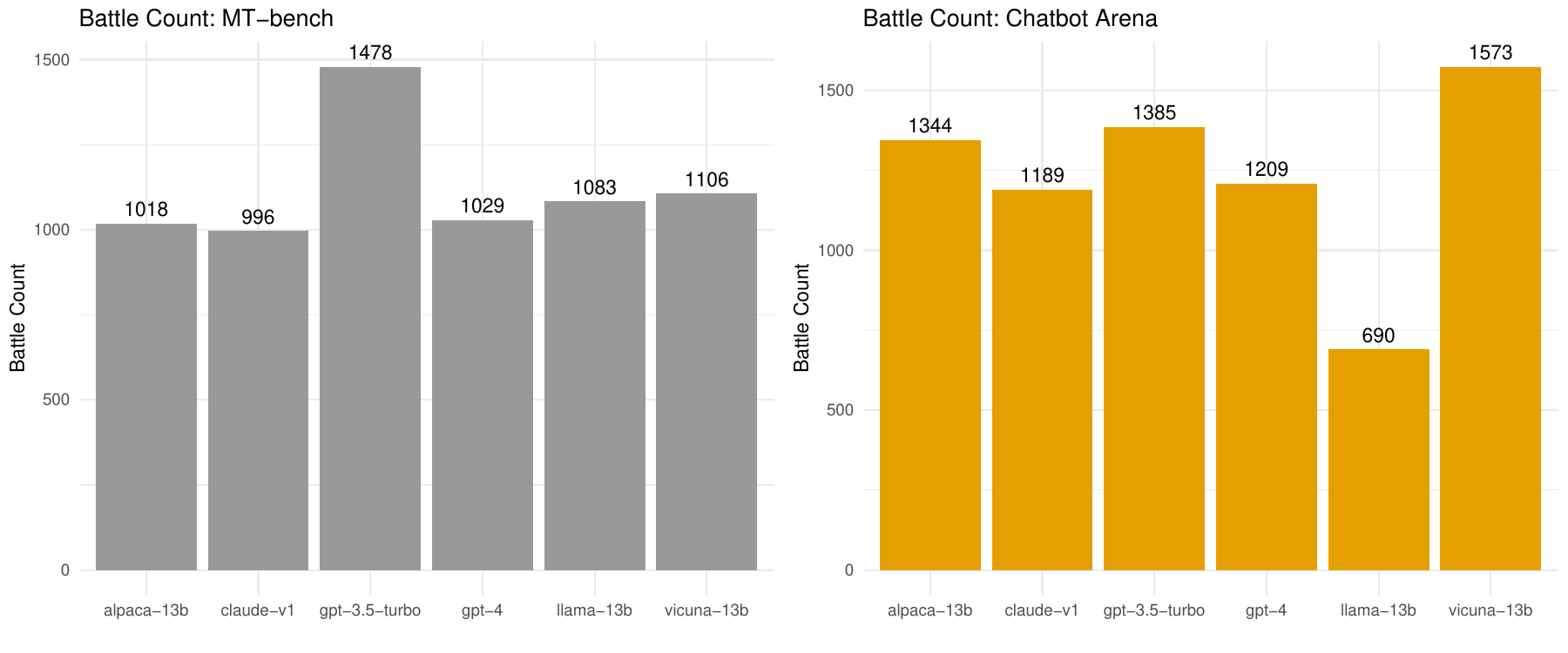} 
    \caption{Battle counts for each model in MT-bench and Chatbot Arena.}
    \label{fig:battle counts}
\end{figure}

\begin{figure}[H]
    \centering
    \includegraphics[width=\linewidth]{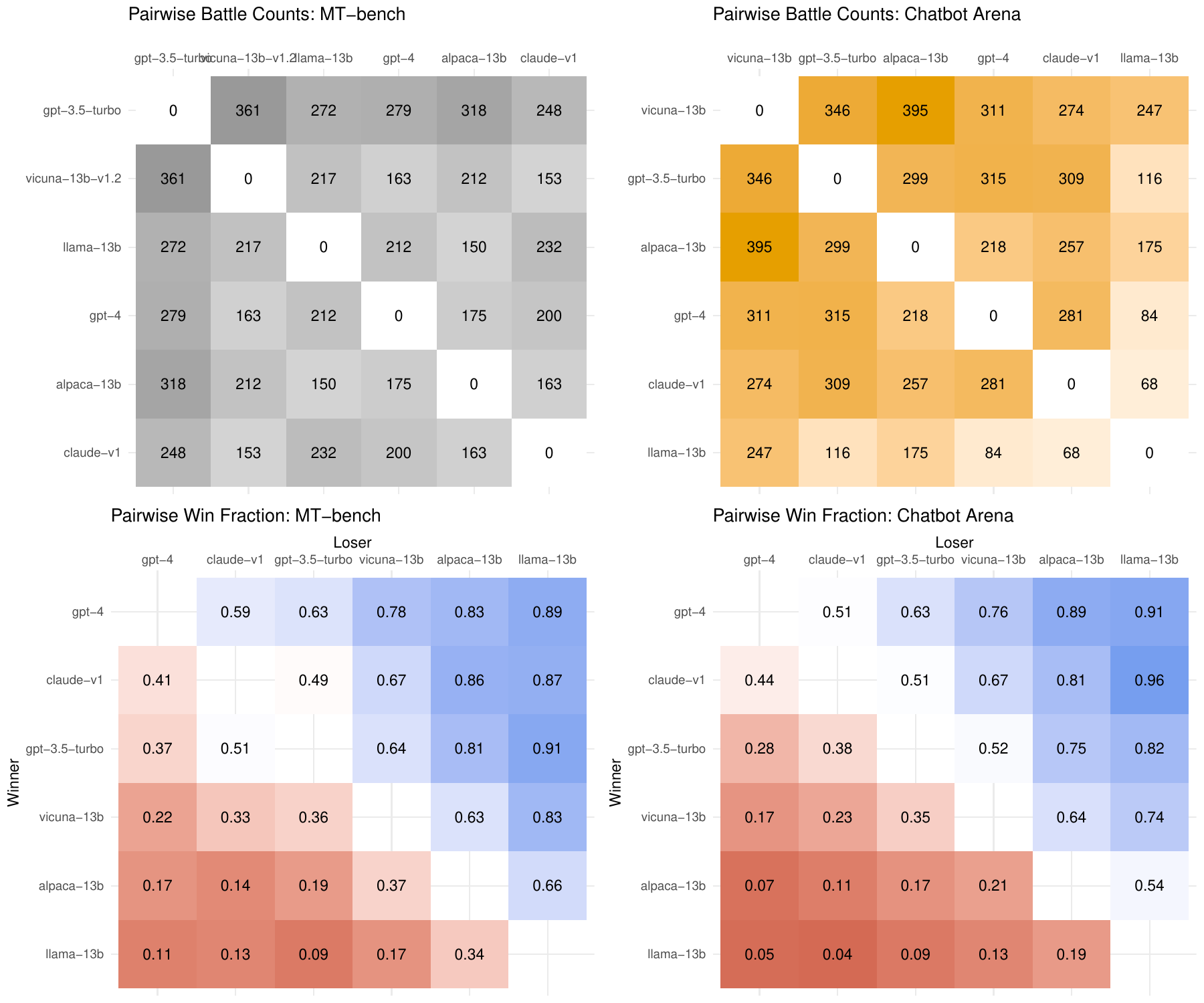}
    \caption{Battle counts and win fractions of each combination of models in MT-bench and Chatbot Arena.}
    \label{fig:battle_combinations}
\end{figure}

\begin{figure}[H]
    \centering
    \includegraphics[width=\linewidth]{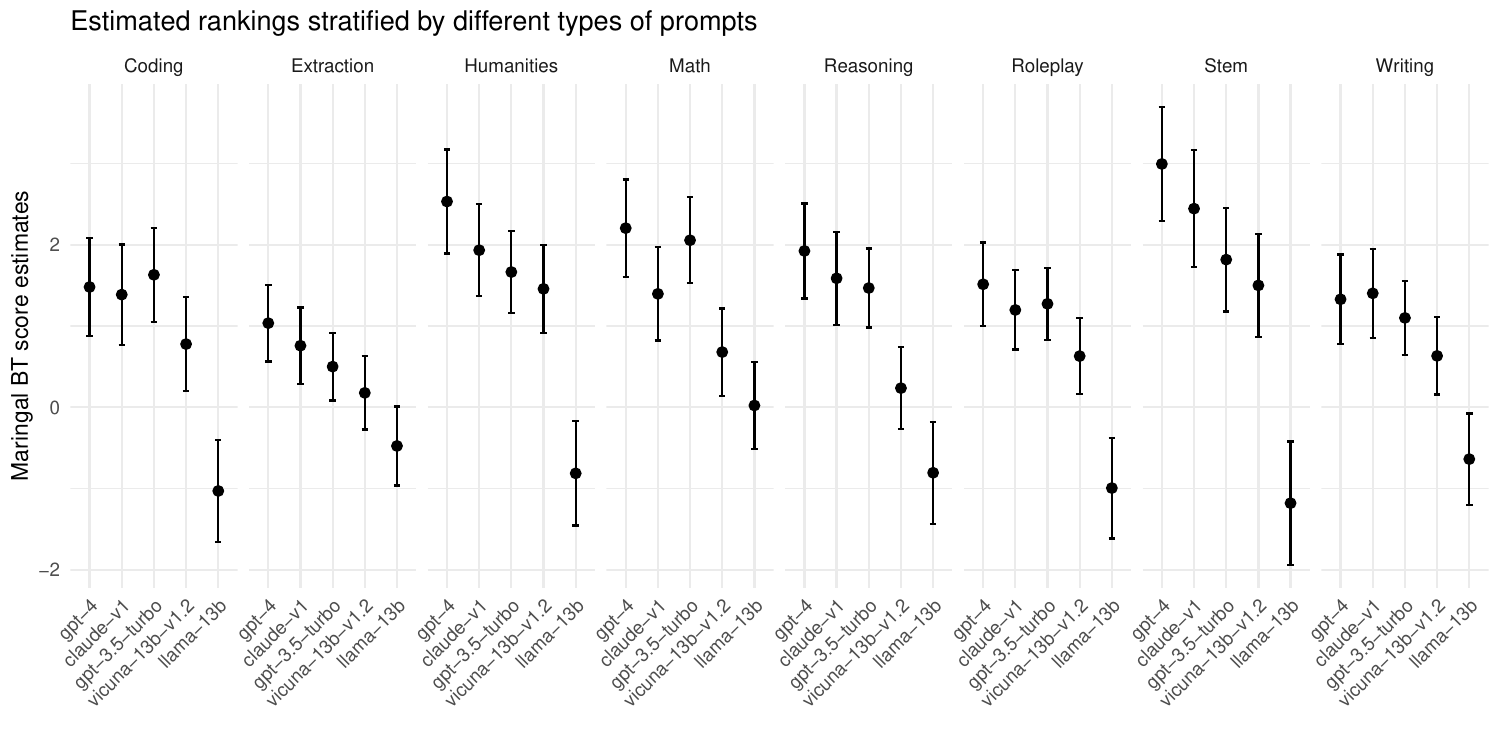}
    \caption{Estimated rankings of each model using marginal BT model with Alpaca-13b as the reference model, stratified by types of prompts.}
    \label{fig:pairwise_wins}
\end{figure}

\begin{table}[H]
\begin{center}
\caption{{Estimated rankings of LLMs using Alpaca-13b as the reference model. Results are presented as estimates, standard errors and 95\% Confidence interval.}\label{tab:main result}}
\begin{adjustbox}{width=\textwidth} 
{
\begin{tabular}{lrrrrrrrrr}
\toprule
   & \multicolumn{3}{c}{$\bphi$} 
& \multicolumn{3}{c}{$\bpsi$} & \multicolumn{3}{c}{Marginal BT} \\
\cmidrule(l){2-4} \cmidrule(l){5-7} \cmidrule{8-10}
  &  Estimate & Std  &  95\% CI   &  Estimate & Std  &  95\% CI &  Estimate & Std  &  95\% CI \\
\midrule
Claude-v1 & 1.54 & 0.09 & (1.36, 1.71) & 1.47 & 0.08 & (1.3, 1.63) & 1.81 & 0.09 & (1.63, 1.98)\\
GPT-3.5-turbo & 1.51 & 0.08 & (1.35, 1.67) & 1.44 & 0.08 & (1.3, 1.59) & 1.36 & 0.08 & (1.2, 1.53)\\
GPT-4 & 1.91 & 0.10 & (1.72, 2.11) & 1.83 & 0.09 & (1.65, 2.01) & 2.10 & 0.09 & (1.92, 2.28)\\
Llama-13b & -0.60 & 0.09 & (-0.78, -0.41) & -0.55 & 0.09 & (-0.72, -0.38) & -0.63 & 0.11 & (-0.85, -0.42)\\
Vicuna-13b-v1.2 & 0.75 & 0.08 & (0.59, 0.91) & 0.70 & 0.08 & (0.55, 0.86) & 0.89 & 0.08 & (0.74, 1.04)\\
\bottomrule
\end{tabular}}
\end{adjustbox}
\end{center}
\end{table}

\end{document}